\documentclass[12pt]{article}
\usepackage[sort&compress,square,comma,numbers]{natbib}
\usepackage{braket,amsmath,amssymb,graphicx,slashed,mathtools}
\usepackage{accents,dsfont}
\usepackage[toc,page]{appendix}
\usepackage[export]{adjustbox}
\pdfoutput=1

\usepackage[
	colorlinks=true,
	citecolor=black,
	linkcolor=black,
	urlcolor=blue,
	hypertexnames=false]{hyperref}

\newcommand{\arXiv}[2]{\href{http://arxiv.org/pdf/#1}{{\tt #2/#1}}}
\newcommand{\arXivold}[1]{\href{http://arxiv.org/pdf/#1}{{\tt #1}}}

\newcommand{\beq}{\begin{equation}}
\newcommand{\eeq}{\end{equation}}

\newcommand{\V}[1]{\mathbf{#1}}
\DeclareMathOperator{\Tr}{Tr}

\begin{document}
\begin{center} 
{\LARGE \bf Tree-Level Factorization Obstruction}\\
\vskip 8pt
{\LARGE \bf in Monopole Production}
\end{center}

\begin{center} 
{\bf Hsing-Yi Lai} and {\bf John Terning}  \\
\end{center}
\vskip 8pt
\begin{center}
{\it Center for Quantum Mathematics and Physics (QMAP),\\Department of Physics, University of California, Davis, CA 95616}
\end{center}

\vspace*{0.1cm}
\begin{center}
{\tt  \href{mailto:hylai@ucdavis.edu}{hylai@ucdavis.edu}\,,
 \href{mailto:jterning@gmail.com}{jterning@gmail.com}}

\end{center}
\centerline{\large\bf Abstract}
\begin{quote}
We show that the tree-level amplitude for producing a monopole--antimonopole pair from an electrically charged pair cannot be constructed from the minimal one-photon couplings. Gluing the electric and magnetic three-point vertices gives a nonzero photon-pole contribution, but the resulting magnetic final state has the opposite discrete-symmetry eigenvalue from the electric initial state. Local four-point interactions cannot change the pole contribution fixed by the three-point couplings, so they cannot repair this mismatch. For fermions, the obstruction is specific to the minimal three-point coupling: an independent Pauli form factor supplies a spin-singlet three-point amplitude and allows the symmetry selection rule to be satisfied.
\end{quote}


\section{Introduction}\label{sec: Introduction}

Computing amplitudes involving both electric and magnetic charges is not straightforward. The Dirac charge quantization condition~\cite{Dirac:1931kp} requires the magnetic coupling to be inversely proportional to the electric coupling, rendering the amplitude nonperturbative. A construction with ``dark'' monopoles coupled to a ``dark'' photon can nevertheless produce a perturbative coupling to the ordinary photon through kinetic mixing~\cite{Terning:2018lsv}. The dark-photon mass $m_D$ confines the monopoles and makes the Dirac string a physical flux tube~\cite{Nielsen:1973cs}. As in QCD, we can separate the hard constituent amplitude from the subsequent flux-tube formation and magnetic ``hadronization.'' We assume a hard scale $Q\gg m_D$, so the production kernel is localized on distances much shorter than the confinement length and serves as a short-distance matching coefficient for the later confining dynamics. Those effects are essential for an observable final state, but they are logically distinct from the question studied here: whether the short-distance tree-level single-photon production amplitude can be constructed. If this scale separation is absent, nonlocal flux-tube dynamics may invalidate an isolated local four-point hard-amplitude description; that regime lies outside our assumptions.

Weinberg~\cite{Weinberg:1965rz} famously showed that the one-photon amplitude for scattering electric and magnetic charges is nonzero but not Lorentz invariant. An all-orders resummation of soft photons exponentiates the Lorentz-violating part into a phase, which becomes a harmless multiple of $2\pi$ when Dirac quantization is imposed~\cite{Terning:2018udc}, and a legitimate function of the confining string direction otherwise. A local description with both electric and magnetic charges requires two constrained vector potentials~\cite{Zwanziger:1970hk,SchwarzSen}, and spurious poles can appear before physical quantities are assembled. Reference~\cite{Terning:2020dzg} showed, after canceling these poles, that the tree-level amplitude for electric particles to produce a magnetic pair vanishes for scalars and fermions:
\beq\label{eq: theprocess}
\includegraphics[height=3cm,valign=c]{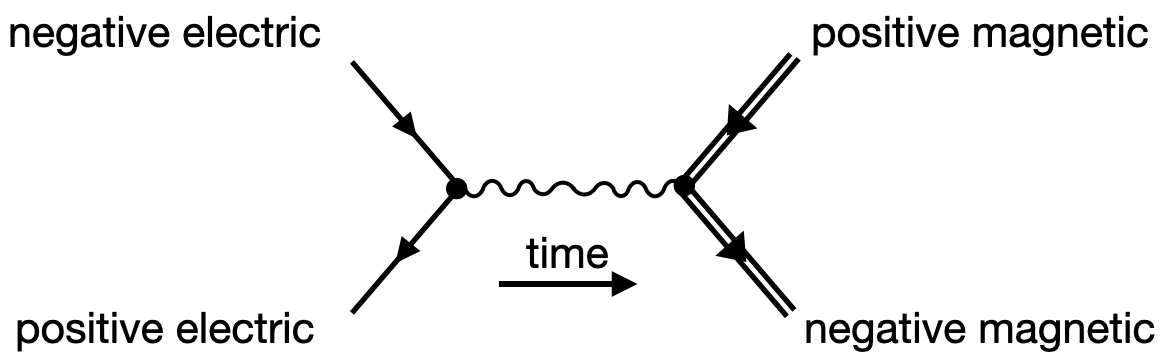}=0~.
\eeq
Here the diagram denotes an electric particle--antiparticle pair producing a magnetic pair through a single virtual photon. The simplicity of the result suggests that its origin should be visible directly in the on-shell state space; the helicity sum that defines the candidate photon residue is discussed in Section~\ref{sec: on-shell}.

Spinor-helicity methods for both massless \cite{Elvang:2013cua} and massive particles~\cite{Arkani-Hamed:2017jhn,Christensen:2022nja,Lai:2023upa} compute amplitudes efficiently without referring to a Lagrangian. Applied to electric--magnetic scattering, they reproduce the result obtained by summing infinitely many soft-photon exchanges~\cite{Csaki:2020inw,Csaki:2022tvb}. This agreement illustrates their usefulness for nonperturbative problems.

Although scattering has been studied with spinor-helicity methods, production is a different four-point problem. In electric--magnetic scattering, the asymptotic state contains both kinds of charge and carries pairwise little-group weight~\cite{Csaki:2020inw,Csaki:2020yei}. In production, the initial state is purely electric and the final state purely magnetic, so the four-point amplitude transforms in the ordinary tensor product of one-particle little groups. The complete scattering amplitude therefore cannot be continued by ordinary crossing into the production state space.

The objects common to the two channels are the three-point vertices, not a four-point amplitude. These vertices carry only the ordinary little-group weights of their three external legs and are well defined independently of pairwise helicity. We now want to determine whether we can use them to construct an ordinary local tree-level production amplitude that preserves the discrete symmetries and contains only the stated spectrum.

\paragraph{Pole and regular terms.}
To see what a local four-point interaction can change, consider the Laurent expansion of a production amplitude near the photon pole,
\beq\label{eq: Laurent symmetry separation}
\mathcal A(s)=\frac{R}{s}+A_0+\mathcal O(s)~.
\eeq
Let $\mathsf G$ be a discrete symmetry of the external states, with required eigenvalue $\eta_G=\pm1$. If $\mathsf G$ leaves $s$ invariant, then
\beq\label{eq: Laurent symmetry action}
\mathsf G\mathcal A-\eta_G\mathcal A
=\frac{\mathsf G R-\eta_G R}{s}
 +\left(\mathsf G A_0-\eta_G A_0\right)+\mathcal O(s)~.
\eeq
Since the powers of $s$ are independent, the pole coefficient and the regular coefficients must satisfy the symmetry condition separately. Thus a regular local term cannot repair a symmetry-forbidden photon-pole coefficient. This is the only analyticity property used in the no-completion argument. In a theory with only electric charges, electric charge conjugation $\mathcal C_E$ and parity $\mathcal P$ are separate discrete symmetries. Including magnetic charges breaks parity but the combined operation of magnetic charge conjugation and parity, $\mathcal C_M\mathcal P$, is a genuine symmetry~\cite{Weinberg:1965rz}. The two symmetries we will use in the production problem do preserve the $s$ channel: the operation $\mathcal C_E\mathcal C_M$ exchanges the particle and antiparticle within each pair, while $\mathcal C_M\mathcal P$ reverses the spatial momenta and conjugates the magnetic charges. In both cases the electric pair remains the initial pair and $(p_1+p_2)^2$ is unchanged, so the symmetry does not map the photon pole into a crossed channel.

There are two different ways to modify the interactions of the theory. We can keep the three-point vertices fixed and add a regular four-point interaction, or we can deform one of the three-point vertices. The first possibility changes $A_0$ and the higher regular coefficients in Eq.~(\ref{eq: Laurent symmetry separation}), but it leaves $R$ unchanged. The second changes the factorization residue itself and can move it into a different symmetry sector. The complete basis below excludes the first possibility for the minimal coupling vertices. For fermions, an independent Pauli form factor provides an explicit example of the second.

\paragraph{Main result.}
We assume for contradiction that an ordinary local tree-level production amplitude $\mathcal A_{\min}$ exists with one photon, the minimal electric and magnetic three-point vertices specified below as its factorization data, ordinary tensor-product production asymptotic states, exact $\mathcal C_M\mathcal P$, and no additional light poles, cuts, or nonlocal string singularities. Define its photon-pole coefficient by
\beq\label{eq: hypothetical minimal residue}
R\equiv \underset{s=0}{\operatorname{Res}}\,\mathcal A_{\min}~.
\eeq
The complete ordinary angular momentum $j=1$ basis and the discrete symmetries require $R=0$. However, ordinary factorization of this same hypothetical amplitude requires
\beq\label{eq: same residue identification intro}
R=\sum_{h=\pm1}\mathcal M_h^E\mathcal M_{-h}^M\equiv R_{\rm req}~,
\eeq
and Section~\ref{sec: on-shell} shows that the helicity sum is generically nonzero. Thus $R_{\rm req}$ is not an independent four-point quantity: it is the value that the residue $R$ must take if $\mathcal A_{\min}$ exists with the stated three-point factorization data. The conditions $R=0$ and $R=R_{\rm req}\neq0$ are incompatible, so we cannot construct the amplitude under these assumptions. Since regular terms cannot change $R$, no finite-derivative local four-point interaction repairs the contradiction.

References~\cite{Ignatiev:1997pm,Terning:2020dzg} establish the two principal prior results. Reference~\cite{Ignatiev:1997pm} derived the discrete-symmetry selection rules, ruling out scalar monopole production, while Ref.~\cite{Terning:2020dzg} found that both the scalar and fermionic production amplitudes vanish in the Zwanziger formulation after the spurious poles are canceled. Those results do not determine whether the vanishing follows from the minimal three-point data, whether an omitted regular four-point interaction can restore the amplitude, or why the same three-point vertices occur in nonzero electric--magnetic scattering amplitudes.

We answer these questions by constructing the complete massive $j=1$ production basis. Since the basis includes every finite-derivative local structure, it excludes a regular four-point completion of the pole coefficient required by the minimal vertices. The physical reason for the fermionic obstruction is particularly simple. A magnetic fermion--antifermion partial wave has
\beq\label{eq: intro spin only selection rule}
\mathcal C_M\mathcal P=(-1)^{S+1}~,
\eeq
where the spin $S=0$, $1$ and the eigenvalue is independent of the orbital angular momentum. Electric--magnetic duality and minimal coupling require the fermionic magnetic pair to be in the spin-triplet sector, and therefore give $\mathcal C_M\mathcal P=+1$, while the electric initial state has eigenvalue $-1$. A scalar magnetic particle--antiparticle pair always has eigenvalue $+1$. For fermions an independent Pauli form factor can change the three-point amplitude to include the spin-singlet sector. The pairwise-covariant scattering analysis then shows why the same minimal vertices can occur in nonzero scattering amplitudes even though they cannot be assembled into the ordinary tree-level production amplitude. Our symmetry analysis is self-contained and uses Ref.~\cite{Ignatiev:1997pm} for comparison rather than as an assumption.

We review constructibility in Section~\ref{sec: constructibility} and the spinor-helicity formalism in Section~\ref{sec: spinor-helicity}, emphasizing angular momentum conservation and minimal coupling. Section~\ref{sec: monopole} reviews the discrete symmetries of magnetic charges. In Section~\ref{sec: CP}, we determine the symmetry-allowed $j=1$ production space and discuss nonminimal Pauli three-point data. Section~\ref{sec: on-shell} computes $R_{\rm req}$ and combines it with the complete basis. We conclude in Section~\ref{sec: conclusion}. The appendices collect spinor-helicity identities, basis details, the all-electric gluing construction, and the corresponding symmetry analysis for electric--magnetic scattering.

\section{Constructibility}\label{sec: constructibility}

On-shell factorization fixes pole residues but does not generally fix independent local contact terms, which appear as boundary contributions in a recursion relation~\cite{Britto:2005fq,ArkaniHamed:2008yf,Cheung:2008dn,Cohen:2010mi,Benincasa:2011pg}. The cited pairwise-covariant constructions realize nonzero electric--magnetic scattering amplitudes using the same minimal three-point vertices. We do not cross that four-point amplitude. Instead, we evaluate the common three-point building blocks on production kinematics and reglue them there, where the asymptotic state belongs to a different little-group representation. The question is whether we can use the resulting production residue to construct a local tree-level amplitude.

\subsection{Why monopole scattering is different}

Electric--magnetic scattering changes the asymptotic little-group representation itself. Electrically and magnetically charged asymptotic states carry a pairwise helicity~\cite{Csaki:2020inw,Csaki:2020yei}
\begin{equation}\label{eq: qij}
h_{ij}=e_i g_j-e_j g_i~,
\end{equation}
where conventional factors of $4\pi$ and $\hbar$ are absorbed into the charge normalization; Dirac quantization therefore makes $h_{ij}$ integer or half-integer in the examples below. The scattering amplitude must transform with the corresponding pairwise little-group phase. Ordinary crossing therefore need not relate this amplitude to a production process whose asymptotic states carry no mutually non-local electric--magnetic pairs.

\subsection{Monopole production}\label{sec: no contact}

For the process in Eq.~(\ref{eq: theprocess}), the initial state contains only electric charges and the final state contains only magnetic charges. The production amplitude therefore transforms in the ordinary tensor product of one-particle little-group representations, so every $j=1$ four-point structure can be classified directly. A local contact term may contain several partial waves, but only its $j=1$ component can complete a single-photon residue. The local-contact completeness argument below shows that this component belongs to the same finite tensor basis as the pole contribution.

For comparison, consider an all-line transverse shift, a complex deformation (as a function of $z$) of all external momenta that preserves on-shellness and momentum conservation. If the shifted amplitude scales as $\mathcal A(z)\sim z^\gamma$ at large $z$, Ref.~\cite{Ema:2024rss} gives
\begin{equation}\label{eq: ALT bound}
\gamma\leq -\frac{N_F}{2}~,
\end{equation}
where $N_F$ is the number of external fermions, for a four-point amplitude with dimensionless couplings. This is consistent with the fermionic result below, but for four scalars it gives only $\gamma\leq0$ and therefore permits a $z^0$ boundary contribution. Recursion alone cannot exclude such a local scalar contact term; this is why we classify the complete $j=1$ contact basis directly. The constructibility estimate is motivational only. It is not used in the complete basis classification, the discrete-symmetry exclusion, or the tree-level obstruction derived below.

\section{Spinor-helicity method}\label{sec: spinor-helicity}

\subsection{Spinor-helicity variables}

The transformation properties of amplitudes are most transparent when expressed in spinor-helicity variables. For null and timelike momenta, they are given by:
\beq\label{eq: massless p}
\text{null: }p_\mu \sigma^\mu_{\alpha\dot{\alpha}} = p_{\alpha\dot{\alpha}}=\lambda_\alpha \Tilde{\lambda}_{\dot{\alpha}}~,\hspace{3mm}
\eeq
\beq\label{eq: massive p}
\text{timelike: }p_{\alpha\dot{\alpha}}=\lambda_\alpha ^I \Tilde{\lambda}_{\dot{\alpha}I}~,~I=1,2~,
\eeq
where $\sigma^{\mu}_{\alpha\dot{\alpha}}=(\mathds{1},\boldsymbol{\sigma})_{\alpha\dot{\alpha}}$ and $\bar{\sigma}^{\mu\dot{\alpha}\alpha}=(\mathds{1},-\boldsymbol{\sigma})^{\dot{\alpha}\alpha}$. Repeated indices, including $\mu$ and $I$, are summed over. Note that for a particle with mass $m$ and momentum $p^\mu$, $\det{p}=\det{\lambda^I} \det{\Tilde{\lambda}_I}=m^2$ and we will take $\det{\lambda^I}=\det{\Tilde{\lambda}_I}=m$. We use the ``West Coast,'' mostly-minus convention $\eta_{\mu\nu}=(+,---)$, which is more suitable for momentum-space calculations. The above definition does not uniquely determine $\lambda$ and $\Tilde{\lambda}$. We may further require them to satisfy the Dirac equation.
\beq
p_{\alpha\dot{\beta}}\Tilde{\lambda}^{I\dot{\beta}}=m\lambda^I_\alpha~,\quad p^{\dot{\alpha}\beta}\lambda^I_{\beta}=m\Tilde{\lambda}^{I\dot{\alpha}}~.
\eeq
These conditions narrow our choices while leaving enough freedom to furnish a representation of the particle's spin.

An amplitude is a Lorentz-invariant expression in which the spinor indices $\alpha$ and $\dot{\alpha}$ are fully contracted using the antisymmetric tensor $\varepsilon_{\alpha\beta}$ and $\varepsilon_{\dot{\alpha}\dot{\beta}}$. Here, $\lambda_\alpha$ transforms as a left-handed spinor and $\Tilde{\lambda}_{\dot{\alpha}}$ transforms as a right-handed spinor. We adopt the convention $\varepsilon^{12}=-\varepsilon_{12}=1$. Such contractions are most conveniently denoted by
\beq
\varepsilon_{\alpha\beta}\lambda^\alpha \chi^\beta = \lambda^\alpha\chi_\alpha=\braket{\lambda\chi}~,\quad\varepsilon^{\dot{\alpha}\dot{\beta}}\Tilde{\lambda}_{\dot{\alpha}}\Tilde{\chi}_{\dot{\beta}}=\Tilde{\lambda}_{\dot\alpha} \Tilde{\chi}^{\dot\alpha}=[\lambda\chi]~.
\eeq
The $I$ index of a massive particle labels a basis vector of its little-group $SU(2)$ representation. We choose a spin basis quantized along the three-momentum; in the corresponding high-energy limit, $I=1,2$ approach positive- and negative-helicity components, respectively. Away from that limit they are basis-dependent spin labels, not Lorentz-invariant helicities. An amplitude involving a particle with spin $S$ is a symmetric tensor carrying $\{I_1,I_2,\cdots,I_{2S}\}$ indices. It transforms in a representation of the little-group $SU(2)$ for each massive particle involved. We refer to these labels as $SU(2)$ indices. For example, a four-point amplitude in which particle 1 is massless with helicity $-3/2$, particle 2 is a massive spin-1/2 fermion, particle 3 is a massive scalar, and particle 4 is a massive vector can be written as
\beq\label{eq: sample amplitude}
\frac{\braket{13^I}\braket{12^J}}{m_2m_3(p_1+p_2)^2}\left(\braket{14^{K_1}}[3_I4^{K_2}]+\braket{14^{K_2}}[3_I4^{K_1}]\right)~.
\eeq
Note how the $SU(2)$ index of particle 3 is fully contracted, so this amplitude transforms as a scalar under its little group\footnote{It does not make sense to contract $SU(2)$ indices of different particles, as their little groups transform independently.}. Throughout most of this paper, we will suppress any uncontracted spinor indices. Since the $SU(2)$ indices will always be symmetrized, we can bold the numbers representing massive particles following \cite{Arkani-Hamed:2017jhn} and suppress the $SU(2)$ indices. The amplitude in Eq.~(\ref{eq: sample amplitude}) can then be written as
\beq
\frac{\braket{1\mathbf{2}}}{m_2m_3(p_1+p_2)^2}\braket{1\mathbf{4}}\bra{1}p_3|\mathbf{4}]~,
\eeq
with the understanding that the $SU(2)$ indices for each particle are separately symmetrized. Some useful properties are summarized in Appendix~\ref{app: formula}.

The conventions used most often below are collected here:
\begin{center}
\begin{tabular}{c|l}
Symbol & Meaning\\
\hline
$|i\rangle,|i]$ & massless spinors for momentum $p_i$\\
$|\mathbf{i}\rangle,|\mathbf{i}]$ & massive spinors with symmetrized little-group $SU(2)$ indices\\
$\langle ij\rangle,[ij]$ & antisymmetric left- and right-handed spinor contractions\\
$x_{ij}$ & massive three-point photon factor, defined in Eq.~(\ref{eq:xfactor})\\
$J^\mu,K^\mu$ & electric and magnetic current four-vectors\\
$\mathcal M_h^E,\mathcal M_h^M$ & electric and magnetic three-point vertex amplitudes for photon helicity $h$\\
$q_{\rm in},q_{\rm out}$ & initial- and final-pair relative momenta\\
$T_{LS}^{\mu}$ & fermion-pair $j=1$ covariant with orbital and spin labels $(L,S)$
\end{tabular}
\end{center}

\subsection{Residue}

If the theory has only three-point interactions at tree level, a four-point tree amplitude has the structure
\beq\label{eq: residues}
\mathcal{M}=\frac{R_s}{s-m_s^2}+\frac{R_t}{t-m_t^2}+\frac{R_u}{u-m_u^2}~,
\eeq
where $s$, $t$, and $u$ are the Mandelstam variables. If we analytically continue the momenta to complex values, $\mathcal{M}$ has poles at $s=m_s^2$, $t=m_t^2$, and $u=m_u^2$. The numerators on the right-hand side of Eq.~(\ref{eq: residues}) are the residues of the corresponding poles. Although these quantities are evaluated at complex momenta, they can be analytically continued to real physical momenta. We therefore refer to the numerators on the right-hand side of Eq.~(\ref{eq: residues}) simply as residues.

Since the Mandelstam variables are Lorentz scalars carrying no little-group weight (nor an $SU(2)$ index), all little-group transformation properties must be carried by the residues. Moreover, upon analytically continuing the momenta to complex values, the residue factorizes at the pole. Take the $s$-pole as an example. When $s\rightarrow m_s^2$, we have $R_s\rightarrow R_L R_R$. Here, $R_L$ and $R_R$ are three-point amplitudes evaluated at the point in complexified momentum space where the internal particle is on shell. We will also refer to these poles as on-shell points.

\subsection{Pairwise spinor-helicity variables}\label{sec: pairwise variables}
When an electric charge $e$ and magnetic charge $g$ are both present in an asymptotic state, this system carries a pairwise angular momentum $eg\hbar$ pointing from one charge to the other, independent of their separation \cite{Thomson}. As a result, the scattering amplitude must transform with an additional $U(1)$ little-group weight \cite{Csaki:2020inw,Csaki:2020yei}. We give only a brief overview here.

Let particle $i$ carry electric charge $e_i$ and magnetic charge $g_i$. Between particle $i$ and $j$, Eq.~(\ref{eq: qij}) gives their pairwise helicity $h_{ij}$. To represent this $U(1)$ little group, we use the momenta $p_i^\mu$ and $p_j^\mu$ to define the ``flat'' null pairwise momenta
\beq
(p_{ij}^{\flat \pm})^\mu=\frac{1}{E_i^c+E_j^c}\left[\pm(E_j^c\pm p_c)p_i^\mu\mp (E_i^c\mp p_c)p_j^\mu\right]~,
\eeq
where $E^c_i$ is the energy of particle $i$ in the center-of-momentum frame, in which both particles' three-momenta have magnitude $p_c$.

Using $p_{ij}^{\flat \pm}$, we define the pairwise spinors~\cite{Csaki:2020inw,Csaki:2020yei}
\beq
\ket{p_{ij}^{\flat +}}~,\quad|p_{ij}^{\flat +}]~,\quad\ket{p_{ij}^{\flat -}}~,\text{ and}\quad|p_{ij}^{\flat -}]~,
\eeq
where $\ket{p_{ij}^{\flat +}}$ and $|p_{ij}^{\flat -}]$ carry the little-group weight associated with the minimal pairwise helicity, opposite to that carried by $\ket{p_{ij}^{\flat -}}$ and $|p_{ij}^{\flat +}]$. These pairwise spinors serve as building blocks in addition to the ordinary spinor-helicity variables for the scattering amplitude.

\subsection{Partial-wave basis}
Using spinor-helicity variables, we can perform a partial-wave expansion in a Lorentz-covariant manner and analyze scattering amplitudes in a basis with definite angular momentum~\cite{Jiang:2020rwz}. Relativistically, angular momentum is captured by the Pauli-Lubanski operator $W^\mu$. Let $P^\mu$ be the total momentum and $M^{\mu\nu}$ the Lorentz generators. Then
\beq
W_\mu=\frac{1}{2}\varepsilon_{\mu\nu\rho\sigma}P^\nu M^{\rho\sigma}~.
\eeq
Let $\mathcal{I}$ and $\mathcal{F}$ denote the sets of initial and final particles, respectively. In a scattering process $\mathcal{I}\rightarrow\mathcal{F}$, the total momentum can be represented by spinor-helicity variables using Eqs.~(\ref{eq: massless p}) and~(\ref{eq: massive p}) by summing over the initial particles only.
\beq
P_{\mathcal{I}}=\sum_{i\in\mathcal{I}}\ket{i}[i|\quad\text{(massless) ,}\quad P_{\mathcal{I}}=\sum_{i\in\mathcal{I}}\ket{i^I}[i_I|\quad\text{(massive).}
\eeq
The Lorentz generators decompose as $M_{\mu\nu}\sigma^\mu_{\alpha\dot{\alpha}}\sigma^\nu_{\beta\dot{\beta}}=\varepsilon_{\alpha\beta}\Tilde{M}_{\dot{\alpha}\dot{\beta}}+M_{\alpha\beta}\varepsilon_{\dot{\alpha}\dot{\beta}}$, where~\cite{Witten:2003nn}
\beq
\begin{aligned}
M_{\mathcal{I},\alpha\beta}^{\text{massless}}=&~i\sum_{i\in\mathcal{I}}\left(\ket{i}_\alpha\frac{\partial}{\partial\bra{i}^{\beta}}+\ket{i}_\beta\frac{\partial}{\partial\bra{i}^{\alpha}}\right)~,\\
\Tilde{M}_{\mathcal{I},\dot{\alpha}\dot{\beta}}^{\text{massless}}=&~i\sum_{i\in\mathcal{I}}\left([i|_{\dot{\alpha}}\frac{\partial}{\partial|i]^{\dot{\beta}}}+[i|_{\dot{\beta}}\frac{\partial}{\partial|i]^{\dot{\alpha}}}\right)~.
\end{aligned}
\eeq
We can extend them to the massive case \cite{Conde:2016izb}.
\beq
\begin{aligned}
M_{\mathcal{I},\alpha\beta}^\text{massive}=&~i\sum_{i\in\mathcal{I}}\left(\ket{i^I}_\alpha\frac{\partial}{\partial\bra{i^I}^{\beta}}+\ket{i^I}_\beta\frac{\partial}{\partial\bra{i^I}^{\alpha}}\right)~,\\
\Tilde{M}_{\mathcal{I},\dot{\alpha}\dot{\beta}}^\text{massive}=&~i\sum_{i\in\mathcal{I}}\left([i_I|_{\dot{\alpha}}\frac{\partial}{\partial|i_I]^{\dot{\beta}}}+[i_I|_{\dot{\beta}}\frac{\partial}{\partial|i_I]^{\dot{\alpha}}}\right)~.
\end{aligned}
\eeq
If electric and magnetic charges are present simultaneously, we must include the contribution of pairwise momenta \cite{Csaki:2020inw,Csaki:2020yei}.
\beq
\begin{aligned}
M_{\mathcal{I},\alpha\beta}^\text{pairwise}=&~i\sum_{i>j,\pm}\left(\ket{p_{ij}^{\flat \pm}}_\alpha\frac{\partial}{\partial\bra{p_{ij}^{\flat \pm}}^\beta}+\ket{p_{ij}^{\flat \pm}}_\beta\frac{\partial}{\partial\bra{p_{ij}^{\flat \pm}}^{\alpha}}\right)~,\\
\Tilde{M}_{\mathcal{I},\dot{\alpha}\dot{\beta}}^\text{pairwise}=&~i\sum_{i>j,\pm}\left([p_{ij}^{\flat \pm}|_{\dot{\alpha}}\frac{\partial}{\partial|p_{ij}^{\flat \pm}]^{\dot{\beta}}}+[p_{ij}^{\flat \pm}|_{\dot{\beta}}\frac{\partial}{\partial|p_{ij}^{\flat \pm}]^{\dot{\alpha}}}\right)~.
\end{aligned}
\eeq
In terms of spinor-helicity variables, the Casimir invariant $W^2=W^\mu W_\mu$ becomes
\beq\label{eq: Pauli Lubanski}
W^2=\frac{P^2_\mathcal{I}}{8}\left(\Tr{\Tilde{M}_\mathcal{I}^2}+\Tr{M_\mathcal{I}^2}\right)-\frac{1}{4}\Tr{\left(P^\intercal_\mathcal{I}M_\mathcal{I}P_\mathcal{I}\Tilde{M}_\mathcal{I}\right)}~,
\eeq
where $(M^2)_{\alpha\beta}\equiv M_\alpha^{~\gamma} M_{\gamma\beta}$, $\Tr{M^2}\equiv (M^2)_\alpha^{~\alpha}$, $\Tr{\Tilde{M}^2}\equiv (\Tilde{M}^2)_{~\dot{\alpha}}^{\dot{\alpha}}$ and the projector $P^\intercal$ is defined by
\beq
P^\intercal_{\mathcal{I}}=\sum_{i\in\mathcal{I}}|i]\bra{i}\quad\text{(massless)~,}\quad P^\intercal_{\mathcal{I}}=\sum_{i\in\mathcal{I}}|i_I]\bra{i^I}\quad\text{(massive).}
\eeq
Some useful relations for the Lorentz generators are:
\beq\label{eq: M on ij}
\begin{aligned}
M_{\mathcal{I}}\braket{\mathbf{ij}}=&0~,\quad\text{if }i, j\in \mathcal{I}\text{ or }i, j\in\mathcal{F}\\
M_{\mathcal{I}}\braket{\mathbf{ij}}=&i\left(\ket{\mathbf{i}}\bra{\mathbf{j}}+\ket{\mathbf{j}}\bra{\mathbf{i}}\right)~,\quad\text{if }i\in \mathcal{I},~j\in\mathcal{F}~.
\end{aligned}
\eeq
The massless version of Eq.~(\ref{eq: M on ij}) follows by removing the boldface particle labels.

Using the Pauli-Lubanski operator and the Casimir invariant Eq.~(\ref{eq: Pauli Lubanski}), we can calculate the total angular momentum $j$ in a scattering process. An amplitude $\mathcal{M}^j$ with definite angular momentum is an eigenfunction satisfying $W^2\mathcal{M}^j=-P_\mathcal{I}^2j(j+1)\mathcal{M}^j$. Moreover, if the process goes through a single-particle $s$ channel, that intermediate particle must have spin $j$ and the residue $R^j$ satisfies $W^2 R^j=-P_\mathcal{I}^2j(j+1)R^j$.

For massless local amplitudes, bases with fixed external content and mass dimension can be generated automatically using the method of Ref.~\cite{Degrande:2025uil}. As a fixed-dimension massless warm-up, Table~\ref{tb: massless Rj} checks the helicity limits of the massive basis; it is not an input to the massive completeness analysis below. All particles are massless, with $\mathcal{I}=\{1,2\}$ and $\mathcal{F}=\{3,4\}$.

\paragraph{Mixed-spin channels.}
The scalar-to-fermion and fermion-to-scalar processes are not physically identical, but their kinematic basis counts are transposes of one another. Rotational invariance therefore allows four independent couplings between the initial $j=1$ state and the four $j=1$ structures in the final-state basis, and likewise four independent couplings in the reverse direction. For scalar-to-fermion production, the initial scalar relative momentum is
contracted with one of the four final-state fermion covariants,
$q_{\rm in}\cdot\widehat T_B$. For the reverse fermion-to-scalar channel,
one instead contracts an initial-state fermion covariant with the final
scalar relative momentum, $T_A\cdot q_{\rm out}$.  Electric--magnetic duality can change the pairwise helicity-dependent coefficients but not these little-group tensors. We therefore display only the scalar-to-fermion orientation and use the transposed basis for the reverse channel.
\begin{table}[htb]
\centering
\begin{tabular}[t]{c|c|c}
particles & $j=0$ & $j=1$\\
\hline
$\text{scalar}\rightarrow\text{scalar}$ & $\braket{12} [12]$ & $\braket{13}[13]-\braket{14}[14]$\\
\hline
$\text{scalar}\rightarrow\text{fermion}$ & \text{none} & $\braket{13}[14]=-\braket{23}[24]$\\
\hline
$\text{fermion}\rightarrow\text{fermion}$ & $[12][34]$ & $[13][24]+[14][23]$\\
 & $\braket{12}\braket{34}$ & $\braket{13}\braket{24}+\braket{14}\braket{23}$\\
 & $[12]\braket{34}$  & $\braket{13}[24]$ \\
 & $\braket{12}[34]$  & $\braket{14}[23]$ \\
 & & $\braket{23}[14]$  \\
 & & $\braket{24}[13]$
\end{tabular}
\caption{Four-point residues for production with mass dimension two and definite angular momentum in the $\mathcal{I}=\{1,2\}$ channel. All particles are massless. Note that the above residues are not for a fixed helicity configuration of the external particles.}
\label{tb: massless Rj}
\end{table}
\subsection{Massive j=1 basis}\label{sec: massive j basis}
For massive particles, we focus only on amplitudes relevant to the single-photon production process, which must have $j=1$ in the $\mathcal{I}=\{1,2\}$ channel, with $m_1=m_2=m$ and $m_3=m_4=m'$. To obtain the complete list of such four-point amplitudes, we start by analyzing the case in which particles 1 and 2 are scalars. Since $p_2^\mu=p_3^\mu+p_4^\mu-p_1^\mu$ by momentum conservation, we can write the amplitude as a function of $p_1$, $p_3$, and $p_4$ only. At tree level, the only possible factorization pole is the $s$-channel pole, with $s=(p_3+p_4)^2$ in our $p_1$, $p_3$, $p_4$ basis. The amplitude may also contain a local contact contribution, which has no pole. Therefore, the factors of $p_1$ can only be in the numerator. The most general four-point amplitude for scalar particles 1 and 2 is
\beq\label{eq: scalar  ansatz}
\mathcal{M} = \sum_{N\geq 0} f_s^N(p_3,p_4)_{\{\alpha,\dot{\alpha}\}}\prod_{i=1}^N p_1^{\dot{\alpha}_i\alpha_i}~,
\eeq
where $f_s^N$ is a function only of $p_3^\mu$ and $p_4^\mu$. The notation $\{\alpha,\dot{\alpha}\}$ denotes all spinor indices contracted with $\prod_{i=1}^N p_1^{\dot{\alpha}_i\alpha_i}$; for $N=0$, this product is defined to be 1. If particles 3 and 4 are fermions, $f_s^N$ will also carry their $SU(2)$ indices, which we will not write out explicitly because they do not affect the following analysis. Note that $\prod_{i=1}^N p_1^{\dot{\alpha}_i\alpha_i}$ is fully symmetric under the exchange of any pair $(\alpha_i,\dot{\alpha}_i)\leftrightarrow(\alpha_j,\dot{\alpha}_j)$, so the asymmetric part of $f_s^N$ under such a swap will not contribute to the amplitude. Thus, we can take $f_s^N$ to be fully symmetric under exchanging any pair of $(\alpha_i,\dot{\alpha}_i)\leftrightarrow(\alpha_j,\dot{\alpha}_j)$ without loss of generality.

For an amplitude to carry definite $j$, it must satisfy
\beq\label{eq: W^2 eigen}
W^2\mathcal{M}^j=-P_{\mathcal{I}}^2 j(j+1)\mathcal{M}^j~.
\eeq
The direct spinor-index evaluation of this condition in the $p_1$ ansatz is not needed for the classification. We therefore use the exact relative-momentum decomposition in the main text; Appendix~\ref{app: scalar detail} gives the corresponding harmonic decomposition. From now on we specialize to $j=1$.

The different orbital components can be separated without leaving the equal-mass shell. Define the relative momentum
\beq\label{eq: relative momentum}
q^\mu\equiv\frac{1}{2}(p_1-p_2)^\mu~,
\qquad
p_1^\mu=\frac{1}{2}(P^T)^\mu+q^\mu~,
\qquad
p_2^\mu=\frac{1}{2}(P^T)^\mu-q^\mu~.
\eeq
Since $p_1^2=p_2^2=m^2$, we have
\beq\label{eq: q constraints}
P^T\cdot q=0~,
\qquad
q^2=m^2-\frac{s}{4}~.
\eeq
We can now introduce the exact complex deformation
\beq\label{eq: exact r deformation}
q(r)=rq~,
\qquad
p_{1,2}(r)=\frac{1}{2}P^T\pm rq~,
\qquad
m_r^2=\frac{s}{4}+r^2\left(m^2-\frac{s}{4}\right)~.
\eeq
For every complex $r$, this deformation keeps $p_1(r)+p_2(r)=P^T$ and satisfies $p_1(r)^2=p_2(r)^2=m_r^2$ exactly. Thus no large-$r$ or approximately equal-mass limit is required.

The Pauli--Lubanski operator rotates the direction of $q$ in the three-dimensional space transverse to $P^T$. A homogeneous polynomial of degree $N$ in $q$ decomposes into transverse symmetric-traceless tensors of ranks $N,N-2,\ldots$. The rank-$\ell$ tensor has orbital angular momentum $\ell$ and therefore satisfies $W^2=-s\ell(\ell+1)$. For scalar particles 1 and 2, $j=1$ selects only the rank-one harmonic. Factors of $q^2=m^2-s/4$ are Lorentz scalars and can be absorbed into its coefficient. Thus the only independent $j=1$ amplitude has the form
\beq\label{eq: scalar final ansatz}
\mathcal{M}^{(j=1)}=(p_1-p_2)^{\dot{\alpha}\alpha}f_s^{(N=1)}(p_3,p_4)_{\alpha\dot{\alpha}}~,
\eeq
Details of this calculation are given in Appendix~\ref{app: scalar detail} for readability. Possible forms of $f_s^{(N=1)}$ depend on the spin of particles 3 and 4. We will discuss them separately, and specify $f^{(N=1)}_{s\rightarrow s}$ (and $f^{(N=1)}_{s\rightarrow f}$) for the all-scalar and scalar-to-fermion cases.
\begin{itemize}
    \item Particles 3 and 4 are scalars.

    In this channel the uniqueness of the $j=1$ structure can be seen directly. In the center-of-momentum frame, the magnitudes of the initial and final relative momenta are fixed once $s$ and the external masses are fixed. A finite-derivative scalar contact amplitude can therefore depend on the scattering angle only through
    \beq\label{eq: scalar angular variable}
    x=\widehat{\mathbf q}_{\rm in}\cdot\widehat{\mathbf q}_{\rm out}~,
    \eeq
    and its angular dependence is a polynomial in $x$. Expanding this polynomial in Legendre polynomials, the $j=1$ projection keeps only $P_1(x)=x$. Thus every scalar contact interaction has
    \beq\label{eq: unique scalar j1 contact}
    \mathcal M_{s\rightarrow s}^{(j=1)}
    =q_{\rm in}\cdot q_{\rm out}\,F_{ss}(s)
    =\frac14 (p_1-p_2)\cdot(p_3-p_4)\,F_{ss}(s)~,
    \eeq
    where the overall sign associated with the center-of-momentum metric convention has been absorbed into $F_{ss}$. Higher powers of $x$ contribute to $j=1$ only through their $P_1(x)$ component, while powers of $q_{\rm in}^2$ and $q_{\rm out}^2$ are radial invariants and only change $F_{ss}(s)$. No independent parity-odd scalar exists because the four external momenta are linearly dependent. Hence $(p_1-p_2)\cdot(p_3-p_4)$ is the unique scalar $j=1$ tensor, not a choice of basis representative.

    Equivalently, $f^{(N=1)}_{s\rightarrow s}(p_3,p_4)_{\alpha\dot{\alpha}}$ must carry its spinor indices through $p_{3\alpha\dot{\alpha}}$ or $p_{4\alpha\dot{\alpha}}$. Since $(p_1-p_2)\cdot p_3=-(p_1-p_2)\cdot p_4$, only their difference is independent, so
    \beq\label{eq: scalar coefficient function}
    f^{(N=1)}_{s\rightarrow s}(p_3,p_4)_{\alpha\dot{\alpha}}
    =(p_3-p_4)_{\alpha\dot{\alpha}}F_{ss}(s)~.
    \eeq
Near the photon pole, a local tree amplitude at fixed order in the derivative expansion has
    \beq\label{eq: scalar Laurent coefficient}
    F_{ss}(s)=\frac{r_{ss}}{s}+P_{ss}(s)~,
    \eeq
where $P_{ss}$ is polynomial. If desired, its coefficients may be normalized by a short-distance scale $\Lambda$ as $P_{ss}(s)=\Lambda^{-2}\sum_{n\geq0}a_n(s/\Lambda^2)^n$. This separates the physical pole from the regular local expansion and does not use an external mass as the EFT normalization scale.

    \item Particles 3 and 4 are fermions.

    To carry one left-handed and one right-handed Lorentz index while transforming as spin-$1/2$ under the little groups of particles 3 and 4, the independent structures may be chosen as $\ket{\mathbf{3}}_\alpha[\mathbf{4}|_{\dot{\alpha}}$, $\ket{\mathbf{4}}_\alpha[\mathbf{3}|_{\dot{\alpha}}$, $\langle\mathbf{34}\rangle p_{3\alpha\dot{\alpha}}$, and $[\mathbf{34}]p_{3\alpha\dot{\alpha}}$. Terms containing $p_4$ are redundant after contraction with $p_1-p_2$. Thus
    \beq\label{eq: f s to f}
    \begin{aligned}
    f^{(N=1)}_{s\rightarrow f}(p_3,p_4)_{\alpha\dot{\alpha}}={}&
    \ket{\mathbf{3}}_\alpha[\mathbf{4}|_{\dot{\alpha}}F_1(s)
    +\ket{\mathbf{4}}_\alpha[\mathbf{3}|_{\dot{\alpha}}F_2(s)\\
    &+p_{3\alpha\dot{\alpha}}\left(\langle\mathbf{34}\rangle F_3(s)+[\mathbf{34}]F_4(s)\right)~.
    \end{aligned}
    \eeq
Each $F_i(s)$ has the mass dimension appropriate to its tensor and, near the photon pole, decomposes uniquely as a pole coefficient divided by $s$ plus a regular polynomial at fixed order in the derivative expansion. Spinor structures such as $\bra{\mathbf{3}}p_4$ reduce to this basis after contraction with $(p_1-p_2)^{\dot{\alpha}\alpha}$.
\end{itemize}
These results and the all-fermion case discussed below are summarized in Table~\ref{tb: massive Mj}.

If particles 1 and 2 are fermions, the amplitude must carry the $SU(2)$ indices of their little groups. In this case, the most general four-point amplitude with only an $s$-channel pole can be written as
\beq\label{eq: fermion ansatz}
\begin{aligned}
    \mathcal{M} = \sum_{N\geq 0}\Big\{&\bra{\mathbf{1}}^{\beta}\bra{\mathbf{2}}^{\gamma}g^N(p_3,p_4)_{\beta\gamma;\{\alpha,\dot{\alpha}\}}+|\mathbf{1}]^{\dot{\beta}}|\mathbf{2}]^{\dot{\gamma}}\bar{g}^N(p_3,p_4)_{\dot{\beta}\dot{\gamma};\{\alpha,\dot{\alpha}\}}\\
    &+\bra{\mathbf{1}}^{\beta}|\mathbf{2}]^{\dot{\beta}}h^N(p_3,p_4)_{\beta\dot{\beta};\{\alpha,\dot{\alpha}\}}+|\mathbf{1}]^{\dot{\beta}}\bra{\mathbf{2}}^{\beta}\bar{h}^N(p_3,p_4)_{\dot{\beta}\beta;\{\alpha,\dot{\alpha}\}}\Big\}\prod_{i=1}^N p_1^{\dot{\alpha}_i\alpha_i}~,
\end{aligned}
\eeq
where $g^N,\bar{g}^N,h^N$, and $\bar{h}^N$ are functions of $p_3^\mu$ and $p_4^\mu$. Terms like $\braket{\mathbf{12}}f$ are included in $\bra{\mathbf{1}}^{\beta}\bra{\mathbf{2}}^{\gamma}g^N_{\beta\gamma;\{\alpha,\dot{\alpha}\}}$, as we can have $g^N_{\beta\gamma;\{\alpha,\dot{\alpha}\}}\propto \varepsilon_{\beta\gamma}$. Below we suppress the arguments of these functions and write $g^N(p_3,p_4)_{\beta\gamma;\{\alpha,\dot{\alpha}\}}=g^N_{\beta\gamma;\{\alpha,\dot{\alpha}\}}$, etc.

Applying the Pauli--Lubanski Casimir to Eq.~(\ref{eq: fermion ansatz}) gives a linear system relating the coefficients at different powers of $p_1$. The full spinor-index expansion is given in Appendix~\ref{app: fermion Casimir}. In the main text we use the equivalent relative-momentum decomposition, which makes the angular-momentum content explicit.

For fermionic particles 1 and 2, the exact deformation in Eq.~(\ref{eq: exact r deformation}) separates the orbital harmonics while preserving the equal-mass shell. Their two-particle spin space decomposes as $\frac{1}{2}\otimes\frac{1}{2}=0\oplus1$. Coupling orbital angular momentum $L$ to spin $S$, a total $j=1$ state can occur only in the four sectors
\beq\label{eq: fermion LS sectors}
(L,S)=(1,0),\quad(0,1),\quad(1,1),\quad(2,1)~.
\eeq
All higher orbital harmonics are therefore absent. These four $(L,S)$ sectors are precisely the four multiplicity labels $A\in\mathfrak A=\{10,01,11,21\}$ used in the covariant counting below; the spinor reduction and the tensors $T_A^\mu$ are two coordinate descriptions of the same four copies of $V_1$. Reducing the covariant sectors with the massive Dirac equations and Schouten identities allows us to choose the following nonredundant representative in the $p_1$-based ansatz of Eq.~(\ref{eq: fermion ansatz})
\beq\label{eq: many h=g=0}
g^{N}_{\beta\gamma}=0~,\quad\bar{g}^{N}_{\dot{\beta}\dot{\gamma}}=0\quad\text{for}\quad N\geq 2~.\quad h^{N}_{\beta\dot{\gamma}}=0~,\quad\bar{h}^N_{\dot{\beta}\gamma}=0\quad \text{for}\quad N\neq 0~.
\eeq
Appendix~\ref{app: fermion detail} gives the exact relative-momentum decomposition and its relation to this spinor basis.

\begin{table}[htb]
\centering
\small
\begin{tabular}{p{0.18\textwidth}|p{0.22\textwidth}|p{0.45\textwidth}}
Description & Notation & Relation to the other descriptions\\
\hline
Massive orbital--spin basis & $T_A^\mu$, $A\in\{10,01,11,21\}$ & Exact finite-mass decomposition into the four $(L,S)=(1,0),(0,1),(1,1),(2,1)$ copies of $V_1$.\\
\hline
Massive spinor basis & $g^N,\bar g^N,h^N,\bar h^N$ & A spinor-coordinate realization of the same four copies. The Dirac and Schouten reductions leave the $N=0,1$ representatives in Eq.~(\ref{eq: fermion j1 reduced basis}).\\
\hline
High-energy helicity coordinates & $a^1_{\lambda_M\lambda_E}(s)$ & The massless opposite-helicity coordinates of the $j=1$ pole data. The fixed minimal vertices select a definite linear combination; these coordinates do not restrict the complete finite-mass contact basis.
\end{tabular}
\caption{Relationship among the covariant, spinor, and helicity descriptions used for a fermion pair. The first two are equivalent descriptions of the complete massive $j=1$ space. The helicity amplitudes are high-energy coordinates for the pole data, not a third complete finite-mass basis and not a restriction on independent local four-point terms.}
\label{tb: fermion basis dictionary}
\end{table}

The direct Casimir reduction is displayed in Appendix~\ref{app: fermion Casimir}. It gives the same four sectors and fixes the redundant $N=0$ and $N=1$ coefficients. We define $g_{\alpha\dot{\alpha}}\varepsilon_{\beta\gamma}\equiv g^{(N=1)}_{\beta\gamma;\alpha\dot{\alpha}}$ and $\bar{g}_{\alpha\dot{\alpha}}\varepsilon_{\dot{\beta}\dot{\gamma}}\equiv \bar{g}^{(N=1)}_{\dot{\beta}\dot{\gamma};\alpha\dot{\alpha}}$. The complete $j=1$ amplitude can then be written as
\beq\label{eq: fermion j1 reduced basis}
\begin{aligned}
    \mathcal{M}^{(j=1)}=&\bra{\mathbf{1}}^{\beta}\bra{\mathbf{2}}^{\gamma}\Bigl(g^{(N=0)}_{\beta\gamma}+g^{(N=0)}_{\gamma\beta}\Bigr)+\braket{\mathbf{12}}\bigl(p_1-p_2\bigr)^{\dot{\alpha}\alpha}g_{\alpha\dot{\alpha}}\\
+&|\mathbf{1}]^{\dot{\beta}}|\mathbf{2}]^{\dot{\gamma}}\Bigl(\bar{g}^{(N=0)}_{\dot{\beta}\dot{\gamma}}+\bar{g}^{(N=0)}_{\dot{\gamma}\dot{\beta}}\Bigr)+[\mathbf{12}]\bigl(p_1-p_2\bigr)^{\dot{\alpha}\alpha}\bar{g}_{\alpha\dot{\alpha}}\\
&+\bra{\mathbf{1}}^\beta|\mathbf{2}]^{\dot{\beta}}h^{(N=0)}_{\beta\dot{\beta}}+|\mathbf{1}]^{\dot{\beta}}\bra{\mathbf{2}}^\beta\bar{h}^{(N=0)}_{\dot{\beta}\beta}~.
\end{aligned}
\eeq
The mixed-chirality coefficients obey the independent constraint
\beq\label{eq: mixed chirality constraint}
(P^T)^{\dot{\beta}\beta}\big(h^{(N=0)}_{\beta\dot{\beta}}-\bar{h}^{(N=0)}_{\dot{\beta}\beta}\big)=0~.
\eeq
Here factors of $1/2$ and an overall sign have been absorbed into the coefficient functions. The first two lines contain the spin-singlet and spin-triplet same-chirality sectors, while the last line contains the mixed-chirality triplet sector. Since the scalar-to-fermion case was already obtained from Eq.~(\ref{eq: f s to f}), we now specialize to the all-fermion basis.

We can now complete the counting without enumerating an open-ended set of spinor substitutions. Let
\beq
q_{\rm out}=\frac{p_3-p_4}{2}~,
\qquad
P=p_3+p_4~,
\qquad
P\cdot q_{\rm out}=0~.
\eeq
For a final fermion pair, define $\widehat T_B^\mu$ by the four covariants in Eq.~(\ref{eq: exact fermion covariants}), with $q$, $\Sigma$, and $S^\mu$ replaced by $q_{\rm out}$, $\widehat\Sigma$, and $\widehat S^\mu$. We use the index set
\beq
\mathfrak A=\{10,01,11,21\}~.
\eeq
The initial and final fermion-pair $j=1$ spaces are therefore
\beq
\mathcal H_{j=1}^{\rm in}\simeq\bigoplus_{A\in\mathfrak A}V_1~,
\qquad
\mathcal H_{j=1}^{\rm out}\simeq\bigoplus_{B\in\mathfrak A}V_1~,
\eeq
where $V_1$ is the three-dimensional spin-one representation. A rotationally invariant amplitude is an intertwiner between these spaces. Rotational invariance allows each of the four initial spin-one structures to couple independently to each of the four final spin-one structures. Since there is only one rotationally invariant contraction between any chosen initial and final spin-one pair, the full amplitude contains $16$ independent tensor structures. In other words, since $\operatorname{Hom}_{SO(3)}(V_1,V_1)$ is one dimensional, Schur's lemma gives
\beq\label{eq: all fermion count}
\dim\operatorname{Hom}_{SO(3)}
\left(\mathcal H_{j=1}^{\rm in},\mathcal H_{j=1}^{\rm out}\right)
=4\times4=16~.
\eeq
A complete all-fermion basis is therefore
\beq\label{eq: all fermion covariant basis}
\mathcal M_{f\to f}^{(j=1)}
=\sum_{A,B\in\mathfrak A}F_{AB}(s)\,
T_A\cdot\widehat T_B~.
\eeq
The independence follows from the commuting orbital- and spin-Casimir operators. The four tensors $T_{10}$, $T_{01}$, $T_{11}$, and $T_{21}$ have distinct eigenvalue pairs
\begin{equation}
(L^2,S^2)=(2,0),(0,2),(2,2),(6,2)~,
\end{equation}
so, wherever they are nonzero, they belong to distinct simultaneous $(L^2,S^2)$ eigenspaces and are linearly independent. In the pair center-of-momentum frame their spatial parts are proportional to $q\,\Sigma$, $S$, $q\times S$, and $q(q\cdot S)-q^2S/3$, respectively, and each is nonzero for generic relative momentum and polarization. The same argument applies to the final tensors $\widehat T_B$. Thus the sixteen products in Eq.~(\ref{eq: all fermion covariant basis}) are independent on the open set $|q_{\rm in}|\,|q_{\rm out}|\neq0$.

\paragraph{Algebraic completeness and threshold kinematics.}
We first establish completeness as a statement about polynomial covariants before assigning numerical momenta. Any polynomial in either relative momentum has the exact trace decomposition
\beq\label{eq: main harmonic decomposition}
q^{\mu_1}\cdots q^{\mu_N}
=\sum_{k=0}^{\lfloor N/2\rfloor}(q^2)^k
q^{\langle\mu_1}\cdots q^{\mu_{N-2k}\rangle}\,\Pi^{(k)}~,
\eeq
where the angular brackets denote the symmetric-traceless projection in the space transverse to $P$, with transverse metric
\beq
\Pi^{\mu\nu}=\eta^{\mu\nu}-\frac{P^\mu P^\nu}{P^2}~,
\qquad P^2=s~,
\eeq
and $\Pi^{(k)}$ denotes products of transverse metrics. The projector is regular at the massive threshold, where $s=4m^2\neq0$, and Eq.~(\ref{eq: main harmonic decomposition}) introduces no inverse powers of $q^2$. Each symmetric-traceless tensor has definite orbital angular momentum $L=N-2k$. Coupling it to $S=0,1$ gives total $j=1$ only in the four sectors $(L,S)=(1,0),(0,1),(1,1),(2,1)$ listed above, while powers of $q^2$ only change the radial coefficient functions. Applying the decomposition separately to $q_{\rm in}$ and $q_{\rm out}$ therefore shows that the tensors in Eq.~(\ref{eq: all fermion covariant basis}) span every local polynomial $j=1$ covariant. This algebraic spanning statement is frame independent and remains valid at threshold.

Pointwise linear independence is a separate question. On the open set $|q_{\rm in}|\,|q_{\rm out}|\neq0$, the distinct $(L^2,S^2)$ eigenvalues make the basis tensors independent and determine the coefficient functions uniquely. At threshold, tensors with $L>0$ vanish after the momenta are evaluated, so the evaluated basis can lose rank and the coefficients need not remain uniquely defined there. These additional relations remove directions from the evaluated span; they cannot create a new polynomial covariant that was absent before evaluation. Thus the basis is algebraically complete on the full polynomial space even though its pointwise rank decreases at special kinematics. This representation-theory count is independent of the spinor Casimir reduction. Appendix~\ref{app: multiplicity basis} displays the full $4\times4$ enumeration and explains its relation to the spinor representatives.

The same counting gives one scalar-to-scalar structure and four scalar-to-fermion structures. The results are summarized in Table~\ref{tb: massive Mj}.

\begin{table}[htb]
\centering
\begin{tabular}[t]{c|c|c}
Particles & Number & Complete $j=1$ basis\\
\hline
$\text{scalar}\to\text{scalar}$
& $1$
& $q_{\rm in}\cdot q_{\rm out}\,F(s)$\\
\hline
$\text{scalar}\to\text{fermion}$
& $4$
& $\displaystyle\sum_{B\in\mathfrak A}F_B(s)\,q_{\rm in}\cdot\widehat T_B$\\
\hline
$\text{fermion}\to\text{fermion}$
& $16$
& $\displaystyle\sum_{A,B\in\mathfrak A}F_{AB}(s)\,T_A\cdot\widehat T_B$
\end{tabular}
\caption{Complete four-point basis with $j=1$ in the $\mathcal I=\{1,2\}$ channel. Here $q_{\rm in}=(p_1-p_2)/2$, $q_{\rm out}=(p_3-p_4)/2$, and $A,B\in\mathfrak A=\{10,01,11,21\}$ label the four fermion-pair multiplicity sectors. The coefficient functions may contain the physical $s$-channel pole. For a local contact term they are polynomials in $s$, so locality changes the coefficient functions but not the tensor basis.}
\label{tb: massive Mj}
\end{table}

\paragraph{Local contact terms.}\label{sec: local contact completeness}
We now show that the same basis contains the $j=1$ component of every finite-derivative local four-point interaction. A local vertex is polynomial in the external momenta and wavefunctions. After projecting onto $j=1$ in the $s$ channel, it lies in the finite representation spaces classified above, with dimensions $1$, $4$, and $16$ for the scalar--scalar, scalar--fermion, and fermion--fermion channels. The tensors in Table~\ref{tb: massive Mj} span these spaces on generic kinematics, and the harmonic decomposition establishes completeness at threshold.

For the all-scalar channel, Eqs.~(\ref{eq: scalar angular variable})--(\ref{eq: unique scalar j1 contact}) give the fixed-$j$ projection explicitly. Every polynomial contact interaction has a unique Legendre expansion, and its $j=1$ component is proportional to $P_1(x)=x$. Thus no scalar contact term can carry $j=1$ without the tensor $(p_1-p_2)\cdot(p_3-p_4)$. This also includes the possible $z^0$ boundary contribution allowed by the all-line shift.

Higher derivatives can add radial powers and higher partial waves, but they cannot enlarge the image of the fixed-$j$ projection. They only change the scalar coefficient functions. At fixed masses, the radial invariants can be written as
\begin{equation}
q_{\rm in}^2=m^2-\frac{s}{4}~,
\qquad
q_{\rm out}^2=m'^2-\frac{s}{4}~,
\end{equation}
so a finite-derivative contact interaction has polynomial coefficients in $s$. An infinite EFT derivative expansion gives the corresponding analytic series order by order. Thus locality changes the coefficients but not the $j=1$ tensor basis.

\subsection{Minimal coupling to the massless photon}\label{sec: minimal coupling}

The amplitude studied below is mediated by the ordinary massless photon. The dark-monopole construction of Ref.~\cite{Terning:2018lsv} supplies a perturbative coupling of this photon to the dark magnetic current through kinetic mixing of the ordinary and dark gauge kinetic terms. The massive dark photon is not the exchanged state in the factorization calculation. Its mass $m_D$ sets the confinement scale of the magnetic sector, so the short-distance production kernel is meaningful in the momentum window
\beq
m_D^2\ll |k^2|\ll \Lambda_M^2~,
\eeq
where the hard process resolves neither the confining flux tube nor the monopole core at scale $\Lambda_M$.

The kinetic-mixing construction fixes the existence and normalization of the perturbative coupling, but it does not by itself remove possible monopole form factors. For a spin-$1/2$ magnetic pair, the current matrix element can be organized as
\beq\label{eq: magnetic current form factors}
\langle M(p_3)\overline M(p_4)|K^\mu(0)|0\rangle
=\overline u(p_3)\left[
\mathcal F_1(k^2)\gamma^\mu
+\frac{i\mathcal F_2(k^2)}{2m_M}\sigma^{\mu\nu}k_\nu
+\cdots
\right]v(p_4)~,
\eeq
where the ellipsis denotes anapole and finite-size structures with additional powers of momentum or inverse powers of the monopole scale. In this paper, minimal coupling means that we retain only the leading Dirac form factor $\mathcal F_1(0)$ at the order being studied. Electric--magnetic duality then acts on the photon field strength, and hence on the photon-helicity phase, without acting on the massive spinor indices. The minimal magnetic vertex therefore has the same $\overline u\gamma^\mu v$ spinor polynomial as the ordinary electric vertex, with the relative helicity phase derived below. The Pauli form factor $\mathcal F_2$ and the remaining finite-size form factors are independent deformations of the three-point data. Thus the small kinetic mixing makes the ordinary-photon coupling perturbative, while the restriction to the Dirac form factor is the substantive minimal-coupling assumption used in the fermion result.

We first derive the helicity restriction imposed by a dimensionless photon coupling. Consider the massless limit of a three-point amplitude with a fermion, an antifermion, and a photon. On the holomorphic branch we can write
\beq\label{eq: holomorphic three point}
\mathcal{A}_3^{\rm hol}=g\langle12\rangle^a\langle23\rangle^b\langle31\rangle^c~.
\eeq
Little-group covariance gives
\beq
 a+c=-2h_1~,
 \qquad a+b=-2h_2~,
 \qquad b+c=-2h_\gamma~,
\eeq
so the total number of angle brackets is
\beq\label{eq: holomorphic degree}
 a+b+c=-(h_1+h_2+h_\gamma)~.
\eeq
A three-point amplitude in four dimensions has mass dimension one. Since a minimal gauge coupling is dimensionless, the spinor polynomial in Eq.~(\ref{eq: holomorphic three point}) must also have dimension one. For a negative-helicity photon, $h_\gamma=-1$, Eq.~(\ref{eq: holomorphic degree}) therefore gives
\beq
 h_1+h_2=0~.
\eeq
The anti-holomorphic branch gives the same result. In that case the total number of square brackets is $h_1+h_2+h_\gamma$, and a positive-helicity photon again requires $h_1+h_2=0$. Thus a dimensionless photon coupling connects a fermion and antifermion of opposite helicity. For example,
\beq\label{eq: massless minimal photon amplitudes}
\mathcal{A}_3^E(1_f^-,2_{\bar f}^+,3_\gamma^-)
 =e\frac{\langle13\rangle^2}{\langle12\rangle}~,
\qquad
\mathcal{A}_3^E(1_f^-,2_{\bar f}^+,3_\gamma^+)
 =e\frac{[23]^2}{[12]}~,
\eeq
with the opposite assignment obtained by exchanging particles~1 and~2. A same-helicity amplitude instead begins with two powers of spinors, for example
\beq
\mathcal{A}_3(1_f^-,2_{\bar f}^-,3_\gamma^-)
 \propto \frac{\langle13\rangle\langle23\rangle}{\Lambda}~,
\eeq
and therefore requires a coefficient of mass dimension $-1$. Such terms require a dimensionful form factor and are nonminimal.

We now show that the same helicity rule holds for a magnetic charge. Let $F_{\mu\nu}^{(h)}$ be the on-shell field strength of a photon with helicity $h=\pm1$, and define
\beq
\widetilde F_{\mu\nu}\equiv\frac{1}{2}\varepsilon_{\mu\nu\rho\sigma}F^{\rho\sigma}~.
\eeq
With our convention for the Levi-Civita tensor, the helicity eigenstates satisfy
\beq\label{eq: dual field helicity eigenvalue}
\widetilde F_{\mu\nu}^{(h)}=-ihF_{\mu\nu}^{(h)}~.
\eeq
An electric--magnetic duality rotation through an angle $\alpha$ acts as
\beq
F_{\mu\nu}\longrightarrow F_{\mu\nu}\cos\alpha+
\widetilde F_{\mu\nu}\sin\alpha~.
\eeq
Using Eq.~(\ref{eq: dual field helicity eigenvalue}), its action on a one-photon helicity state is therefore
\beq\label{eq: duality phase helicity}
F_{\mu\nu}^{(h)}\longrightarrow e^{-ih\alpha}F_{\mu\nu}^{(h)}~.
\eeq
A pure magnetic source is obtained from a pure electric source by the duality rotation $\alpha=\pi/2$, up to the normalization of its charge. Hence its minimal three-point amplitude is
\beq\label{eq: electric magnetic three point relation}
\mathcal{M}_h^M=e^{-ih\pi/2}\frac{g_M}{e}\mathcal{M}_h^E~.
\eeq
For the two photon helicities,
\beq
 e^{-ih\pi/2}=\begin{cases}
 -i~, & h=+1~,\\
 +i~, & h=-1~.
 \end{cases}
\eeq
We now choose one common phase convention for the magnetic external states, $\mathcal M_h^M\rightarrow(-i)\mathcal M_h^M$. This operation is independent of $h$, while
\beq
 (-i)e^{-ih\pi/2}=\begin{cases}
 -1~, & h=+1~,\\
 +1~, & h=-1~,
 \end{cases}
 =-h~.
\eeq
With this convention, Eq.~(\ref{eq: electric magnetic three point relation}) becomes
\beq\label{eq: magnetic helicity phase rule}
\mathcal{M}_h^M=-h\frac{g_M}{e}\mathcal{M}_h^E~.
\eeq
The common phase is conventional, but the relative sign between the two photon helicities cannot be removed. A rephasing-invariant way to state the result is
\beq\label{eq: duality phase invariant}
\Xi\equiv
\frac{\mathcal M_+^M/\mathcal M_+^E}
     {\mathcal M_-^M/\mathcal M_-^E}=-1~.
\eeq
Any common redefinition of the electric states, magnetic states, photon states, or overall magnetic vertex cancels in $\Xi$. Superscripts $E$ and $M$ label the electric and magnetic three-point vertices, while the subscript $h$ labels photon helicity; the corresponding current four-vectors are denoted by $J^\mu$ and $K^\mu$, and the magnetic coupling by $g_M$.
Thus duality changes only the phase associated with the photon helicity. It does not act on the massive spinor indices and therefore cannot change the fermion-helicity selection. In massive spinor-helicity variables, the electric and magnetic minimal vertices can be written as
\beq\label{eq: electric magnetic minimal fermion vertices}
\begin{aligned}
\mathcal{M}_{-}^E&=e\frac{[\mathbf{12}]}{x_{12}}~,
&\qquad \mathcal{M}_{+}^E&=e x_{12}\langle\mathbf{12}\rangle~,\\
\mathcal{M}_{-}^M&=g_M\frac{[\mathbf{12}]}{x_{12}}~,
&\qquad \mathcal{M}_{+}^M&=-g_M x_{12}\langle\mathbf{12}\rangle~.
\end{aligned}
\eeq
The magnetic vertex has exactly the same spinor polynomials as the electric vertex, with the relative sign required by Eq.~(\ref{eq: magnetic helicity phase rule}). To take the high-energy limit, choose the massive little-group components that become the massless helicity spinors. Then $[\mathbf{12}]$ and $\langle\mathbf{12}\rangle$ reduce to $[12]$ and $\langle12\rangle$, while the $x_{12}^{\pm1}$ factor supplies the remaining photon little-group weight. The two expressions therefore reduce to the opposite-helicity amplitudes in Eq.~(\ref{eq: massless minimal photon amplitudes}). This is a useful massless check, but it does not by itself determine the exact finite-mass pole subspace.

\paragraph{Finite-mass spin content and a transverse representative.}
The exact finite-mass statement needed for the symmetry proof concerns the two-particle spin, not a unique orbital continuation of the photon helicity phase. For a fermion pair of mass $m$, define the vector current
\beq\label{eq: exact vector axial currents}
V_m^\mu=\overline u(p_1)\gamma^\mu v(p_2)~.
\eeq
Appendix~\ref{app: exact minimal pole} gives the direct finite-mass reduction. Up to an overall normalization convention,
\beq\label{eq: exact vector current LS}
V_m^\mu
=a_m(s)T_{01}^\mu+b_m(s)T_{21}^\mu~,
\qquad
a_m(s)=\frac{2}{3}(\sqrt{s}+m)~,
\qquad
b_m(s)=\frac{4}{\sqrt{s}+2m}~.
\eeq
The coefficients have different mass dimensions because $T_{21}$ contains two more powers of relative momentum than $T_{01}$. Thus the dimension-four vector vertex creates a fixed ${}^3S_1$--${}^3D_1$ combination and is entirely spin triplet. The minimal magnetic vertex contains the same massive spinor polynomial as the electric vertex; duality supplies only the invariant relative photon-helicity phase in Eq.~(\ref{eq: duality phase invariant}). It therefore does not change the pair spin:
\beq\label{eq: exact minimal magnetic spin sector}
S_M=1~.
\eeq
This exact statement is sufficient for the discrete-symmetry argument because every fermion--antifermion $j=1$ state with $S=1$ has $\mathcal C_M\mathcal P=+1$, independent of $L$.

For intuition, away from the singular factorization limit one may represent the relative helicity phase on the two physical transverse photon polarizations by the transverse axial current
\beq\label{eq: exact axial current LS}
A_{m\perp}^\mu
=\left(\eta^{\mu\nu}-\frac{P^\mu P^\nu}{s}\right)
\overline u(p_1)\gamma_\nu\gamma^5v(p_2)
\propto T_{11}^\mu~.
\eeq
For circular polarization $h=\pm1$ in a timelike pair rest frame,
\beq\label{eq: axial vector helicity relation}
\epsilon_h\cdot A_{m\perp}
=h\,\beta_m\,\epsilon_h\cdot V_m~,
\qquad
\beta_m=\sqrt{1-\frac{4m^2}{s}}~,
\eeq
up to a common phase convention. This relation fixes only the two transverse helicity matrix elements. It does not determine the longitudinal $m_j=0$ component of a full massive $SO(3)$ vector, and the rest-frame parametrization is singular at the photon factorization locus $s=0$. Accordingly, $T_{11}={}^3P_1$ is used only as a convenient transverse representative of the helicity phase, not as a uniquely defined exact finite-mass pole direction. The production argument below uses only Eq.~(\ref{eq: exact minimal magnetic spin sector}).

For comparison, the massless limit resolves a general $j=1$ pole into the opposite-helicity coordinates shown in Table~\ref{tb: minimal-coupling}. We use these helicity coordinates only to display the pole data selected by the minimal three-point vertices; every claim of local completeness refers to the exact finite-mass basis in Table~\ref{tb: massive Mj}. The fixed electric and magnetic phase relations select a definite linear combination of the high-energy coordinates.

\begin{table}[htb]
\centering
\begin{tabular}[t]{c|c|c}
Particles & Number & High-energy opposite-helicity coordinates\\
\hline
$\text{scalar}\to\text{scalar}$
& $1$
& $q_{\rm in}\cdot q_{\rm out}\,F(s)$\\
\hline
$\text{scalar}\to\text{fermion}$
& $2$
& $a^1_{+0}(s),\ a^1_{-0}(s)$\\
\hline
$\text{fermion}\to\text{fermion}$
& $4$
& $a^1_{\lambda_M\lambda_E}(s)$, $\lambda_E,\lambda_M=\pm1$
\end{tabular}
\caption{High-energy helicity coordinates for a $j=1$ pole. The fixed minimal three-point vertices select a definite combination of these entries. This table is not an exact finite-mass restriction on the complete local four-point basis in Table~\ref{tb: massive Mj}.}
\label{tb: minimal-coupling}
\end{table}

The parity transformation of the massive spinors used below is summarized by $\langle\,\rangle\leftrightarrow[\,]$ together with helicity reversal. A derivation, including the signs from contracted $SU(2)$ indices, is given in Appendix~\ref{app: parity}.

\section{Magnetic monopole}\label{sec: monopole}

Weinberg showed that there are two inequivalent helicity phases with which a photon can couple to a conserved current \cite{Weinberg:1965rz}. To compare opposite photon helicities without equating objects of different little-group weight, write each three-point amplitude as a helicity-dependent kinematic tensor times a reduced coefficient $c_h$. For an electric current $J^\mu$ the reduced coefficients are in phase, whereas for a magnetic current $K^\mu$ they have the opposite relative phase. Up to independent overall electric and magnetic normalizations,
\beq\label{eq: magnetic interaction}
c_+^E=c_-^E~,
\qquad
c_+^M=-c_-^M~.
\eeq
Under parity $\mathcal P$, the corresponding combination of polarization tensors $(\epsilon_+^\mu-\epsilon_-^\mu)$ picks up an additional minus sign~\cite{Weinberg:1965rz}, so the magnetic interaction is not parity invariant. The derivation in Section~\ref{sec: minimal coupling} shows that this helicity-dependent phase is the only difference between the minimal electric and magnetic three-point amplitudes: electric--magnetic duality acts on the photon helicity state but leaves the massive spinor polynomial unchanged. This distinct behavior under discrete symmetries has been shown to strongly constrain the production amplitude from electric to magnetic charges \cite{Ignatiev:1997pm}. We will revisit these constraints in Section~\ref{sec: CP} using spinor-helicity variables.

Although $\mathcal{P}$ is broken in the presence of magnetic monopoles, new discrete symmetries emerge. We may define a magnetic charge conjugation operator $\mathcal{C}_M$, which exchanges north and south poles while acting trivially on electric charges and the photon field. Under $\mathcal{C}_M$, the magnetic current $K^\mu$ acquires a minus sign, whereas the electric current $J^\mu$ remains unchanged. The minus sign of $K^\mu$ under $\mathcal{C}_M$ cancels the additional minus sign from the photon field under $\mathcal{P}$. Consequently, both types of interactions are invariant under the combined operation $\mathcal{C}_M\mathcal{P}$. Thus the $\mathcal{C}_M\mathcal{P}$ symmetry should be preserved even though $\mathcal{P}$ itself is broken by magnetic monopoles\footnote{The transformation denoted $PM$ in Ref.~\cite{Ignatiev:1997pm} corresponds to our $\mathcal{C}_M\mathcal{P}$.}. Let $\mathcal{C}_E$ denote the conventional charge conjugation operator that exchanges positive and negative electric charges, sends the photon field to minus itself, and acts trivially on magnetic charges. In the presence of magnetic monopoles, the combined operation $\mathcal{C}_E\mathcal{C}_M$ remains a valid symmetry of the system. Overall, the symmetries $\mathcal{C}_M\mathcal{P}$ and $\mathcal{C}_E\mathcal{C}_M$ persist and constrain the structure of scattering amplitudes\footnote{For ``dark'' electric and magnetic charges, they transform just like regular electric/magnetic charges under $\mathcal{C}_M\mathcal{P}$ and $\mathcal{C}_E\mathcal{C}_M$, so that the regular kinetic mixing is invariant under these transformations.}.

To make the discussion precise, we introduce the following notation. Let $a_p^{h\dagger}$ denote the creation operator for a negatively charged electric particle with momentum $p^\mu$ and helicity $h$. The corresponding antiparticle, carrying positive electric charge, is created by $b^{h\dagger}_p$. A positively charged magnetic monopole is created by $\alpha_p^{h\dagger}$ and its corresponding antiparticle, carrying negative magnetic charge, is created by $\beta_p^{h\dagger}$. We choose their relative phase under magnetic charge conjugation so that
\beq\label{eq: CM creation operator convention}
\mathcal C_M\alpha_p^{h\dagger}\mathcal C_M^{-1}=\beta_p^{h\dagger}~,
\qquad
\mathcal C_M\beta_p^{h\dagger}\mathcal C_M^{-1}=\alpha_p^{h\dagger}~,
\eeq
with $\mathcal C_M^2=1$. More general reciprocal phases cancel in a monopole--antimonopole state and do not change the partial-wave eigenvalues below. For magnetic-pair production in Eq.~(\ref{eq: theprocess}), the $S$-matrix element is
\beq
\bra{0}\alpha_3^{h_3}\beta_4^{h_4}\hat{S}\,a_1^{h_1\dagger}b_2^{h_2\dagger}\ket{0}=i(2\pi)^4 \delta^4(p_1^\mu+p_2^\mu-p_3^\mu-p_4^\mu)\mathcal{M}~.\label{eq: production S matrix}
\eeq
Here we abbreviate $a_1^h\equiv a_{p_1}^h$. We denote the all-scalar production amplitude by $\mathcal{M}_{s\rightarrow s}(1,2;3,4)$, where each number labels the corresponding particle momentum. We denote the scalar-to-fermion and all-fermion amplitudes by $\mathcal{M}_{s\rightarrow f}(1,2;3,4)_{\{h\}}$ and $\mathcal{M}_{f\rightarrow f}(1,2;3,4)_{\{h\}}$, respectively, where $\{h\}$ specifies the fermion helicities. A bar over a particle label denotes its parity-conjugated momentum, and $\{h\}\rightarrow\{-h\}$ denotes the simultaneous helicity reversal of all particles.

Since our process is symmetric under $\mathcal{C}_M\mathcal{P}$, we have $(\mathcal{C}_M\mathcal{P})^{-1}\hat S\mathcal{C}_M\mathcal{P}=\hat S$. We apply the symmetry directly to the production matrix element in Eq.~(\ref{eq: production S matrix}), keeping the initial state on the right of $\hat S$ and the final state on the left throughout.

For scalar-to-scalar production,
\beq
\begin{aligned}
\bra{0}\alpha_3\beta_4\hat S\,a_1^\dagger b_2^\dagger\ket{0}
&=\bra{0}\beta_{\bar3}\alpha_{\bar4}\hat S\,a_{\bar1}^\dagger b_{\bar2}^\dagger\ket{0}\\
&=\bra{0}\alpha_{\bar4}\beta_{\bar3}\hat S\,a_{\bar1}^\dagger b_{\bar2}^\dagger\ket{0}~,
\end{aligned}
\eeq
where the scalar operators commute. Hence
\beq
\mathcal M_{s\rightarrow s}(1,2;3,4)
=+\mathcal M_{s\rightarrow s}(\bar1,\bar2;\bar4,\bar3)~.
\eeq
Applying $\mathcal C_E\mathcal C_M$ instead gives
\beq
\begin{aligned}
\bra{0}\alpha_3\beta_4\hat S\,a_1^\dagger b_2^\dagger\ket{0}
&=\bra{0}\beta_3\alpha_4\hat S\,b_1^\dagger a_2^\dagger\ket{0}\\
&=\bra{0}\alpha_4\beta_3\hat S\,a_2^\dagger b_1^\dagger\ket{0}~,
\end{aligned}
\eeq
and therefore
\beq
\mathcal M_{s\rightarrow s}(1,2;3,4)
=+\mathcal M_{s\rightarrow s}(2,1;4,3)~.
\eeq

For scalar-to-fermion production, the odd intrinsic parity of the final fermion--antifermion pair and the exchange of the two final fermion operators each contribute a minus sign under $\mathcal C_M\mathcal P$, so they cancel:
\beq
\mathcal M_{s\rightarrow f}(1,2;3,4)_{\{h\}}
=+\mathcal M_{s\rightarrow f}(\bar1,\bar2;\bar4,\bar3)_{\{-h\}}~.
\eeq
Under $\mathcal C_E\mathcal C_M$, only the final fermion exchange contributes, giving
\beq
\mathcal M_{s\rightarrow f}(1,2;3,4)_{\{h\}}
=-\mathcal M_{s\rightarrow f}(2,1;4,3)_{\{h\}}~.
\eeq

For fermion-to-fermion production, the two odd pair intrinsic parities cancel under $\mathcal C_M\mathcal P$, while reordering the final magnetic fermions contributes one minus sign. Thus
\beq
\mathcal M_{f\rightarrow f}(1,2;3,4)_{\{h\}}
=-\mathcal M_{f\rightarrow f}(\bar1,\bar2;\bar4,\bar3)_{\{-h\}}~.
\eeq
Under $\mathcal C_E\mathcal C_M$, the electric-pair and magnetic-pair exchanges each contribute a minus sign, and hence
\beq
\mathcal M_{f\rightarrow f}(1,2;3,4)_{\{h\}}
=+\mathcal M_{f\rightarrow f}(2,1;4,3)_{\{h\}}~.
\eeq
The results are collected in Table~\ref{tb: CP of amp}.
\begin{table}[htb]
\centering
\begin{tabular}[t]{|c|l|}
\hline
$\mathcal{C}_M\mathcal{P}$ & $\mathcal{M}_{s\rightarrow s}(1,2;3,4)=+\mathcal{M}_{s\rightarrow s}(\bar{1},\bar{2};\bar{4},\bar{3})$ \\
& $\mathcal{M}_{s\rightarrow f}(1,2;3,4)_{\{h\}}=+\mathcal{M}_{s\rightarrow f}(\bar{1},\bar{2};\bar{4},\bar{3})_{\{-h\}}$\\
& $\mathcal{M}_{f\rightarrow f}(1,2;3,4)_{\{h\}}=-\mathcal{M}_{f\rightarrow f}(\bar{1},\bar{2};\bar{4},\bar{3})_{\{-h\}}$\\
\hline
$\mathcal{C}_E\mathcal{C}_M$ & $\mathcal{M}_{s\rightarrow s}(1,2;3,4)=+\mathcal{M}_{s\rightarrow s}(2,1;4,3)$\\
& $\mathcal{M}_{s\rightarrow f}(1,2;3,4)_{\{h\}}=-\mathcal{M}_{s\rightarrow f}(2,1;4,3)_{\{h\}}$\\
& $\mathcal{M}_{f\rightarrow f}(1,2;3,4)_{\{h\}}=+\mathcal{M}_{f\rightarrow f}(2,1;4,3)_{\{h\}}$\\
\hline
\end{tabular}
\caption{Transformation properties under $\mathcal{C}_M\mathcal{P}$ and $\mathcal{C}_E\mathcal{C}_M$ for electric-pair-to-magnetic-pair production amplitudes. The fermion-to-scalar channel is omitted using the duality redundancy stated in Section~\ref{sec: spinor-helicity}.}
\label{tb: CP of amp}
\end{table}

\section{Vanishing of the $j=1$ production amplitude}\label{sec: CP}
We now determine which $j=1$ production states are compatible with the discrete symmetries. Since a single photon carries $j=1$, only the $j=1$ component of a four-point amplitude can contain its pole. Section~\ref{sec: local contact completeness} shows that the pole and regular terms use the same complete tensor basis. For scalars, the symmetry excludes the only available basis tensor. For fermions, minimal coupling fixes the magnetic pair to the spin-triplet sector in Eq.~(\ref{eq: exact minimal magnetic spin sector}), and we find that this sector has the wrong $\mathcal C_M\mathcal P$ eigenvalue. Equations~(\ref{eq: Laurent symmetry separation}) and~(\ref{eq: Laurent symmetry action}) then convert these selection rules into the production constraint used in Section~\ref{sec: on-shell}.

\subsection{Scalar-to-scalar production}\label{sec: all-scalar}
Equations~(\ref{eq: scalar angular variable})--(\ref{eq: unique scalar j1 contact}) show directly that the $j=1$ projection of any finite-derivative scalar contact interaction is proportional to a single tensor. In the complete basis of Table~\ref{tb: massive Mj}, this tensor is
\beq\label{eq: scalar j1 production basis}
B_{ss}=(p_1-p_2)\cdot(p_3-p_4)~.
\eeq
Near the photon pole, the most general local tree amplitude at fixed order in the derivative expansion is
\beq\label{eq: scalar local contact term}
\mathcal M_{s\rightarrow s}^{(j=1)}
=B_{ss}\left(\frac{r_{ss}}{s}+P_{ss}(s)\right)~,
\eeq
where $P_{ss}(s)$ is polynomial. The coefficient $r_{ss}$ is the unique Laurent residue, while $P_{ss}$ contains the regular local four-point operators, including the constant boundary contribution allowed by the all-line estimate.

Under $\mathcal C_M\mathcal P$, $s$ is unchanged but
\beq
B_{ss}(\bar1,\bar2;\bar4,\bar3)
=(\bar p_1-\bar p_2)\cdot(\bar p_4-\bar p_3)
=-B_{ss}(1,2;3,4)~.
\eeq
This conflicts with the required even transformation in Table~\ref{tb: CP of amp}. Applying Eq.~(\ref{eq: Laurent symmetry action}) sets $r_{ss}=0$ and also sets every coefficient in $P_{ss}$ to zero. The structure satisfies $\mathcal C_E\mathcal C_M$, but it cannot satisfy $\mathcal C_M\mathcal P$. Thus the scalar $j=1$ photon residue and every regular local four-point term vanish at tree level, in agreement with Ref.~\cite{Ignatiev:1997pm}. The possible $z^0$ boundary contribution is the constant term of $P_{ss}$ and is excluded without using the large-$z$ estimate.

\subsection{Fermionic final states}

The fermionic obstruction follows directly from the spin of the magnetic pair. Consider a monopole--antimonopole partial wave $\ket{(M\bar M)LS;JM}$, with the relative momentum and spin phases fixed by Eq.~(\ref{eq: CM creation operator convention}). Under parity, the relative momentum reverses and contributes $(-1)^L$. The intrinsic parities of a Dirac particle and antiparticle have product $-1$, so
\beq\label{eq: fermion pair parity eigenvalue}
\mathcal P\ket{(M\bar M)LS;JM}
=(-1)^{L+1}\ket{(M\bar M)LS;JM}~.
\eeq
Under $\mathcal C_M$, exchanging the monopole and antimonopole reverses the relative momentum, giving $(-1)^L$. Interchanging the two coupled spinors gives $(-1)^{S+1}$, while restoring the fermionic creation operators to their standard order gives one additional minus sign. Thus
\beq\label{eq: fermion pair charge conjugation eigenvalue}
\mathcal C_M\ket{(M\bar M)LS;JM}
=(-1)^L(-1)^{S+1}(-1)\ket{(M\bar M)LS;JM}
=(-1)^{L+S}\ket{(M\bar M)LS;JM}~.
\eeq
Combining Eqs.~(\ref{eq: fermion pair parity eigenvalue}) and~(\ref{eq: fermion pair charge conjugation eigenvalue}) gives
\beq\label{eq: fermion pair PC eigenvalues}
\boxed{\mathcal C_M\mathcal P=(-1)^{S+1}}~.
\eeq
The intrinsic-parity and operator-reordering signs are the same ones used in deriving the production-amplitude transformations collected in Table~\ref{tb: CP of amp}. Equation~(\ref{eq: fermion pair PC eigenvalues}) is therefore the partial-wave form of those conventions. Its important consequence is that the combined eigenvalue is independent of $L$.

The minimal magnetic three-point vertex has the same massive spinor polynomial as the electric vector vertex, and the duality phase acts only on photon helicity. Equation~(\ref{eq: exact minimal magnetic spin sector}) therefore gives
\beq\label{eq: minimal magnetic CP sector}
S_M=1~,
\qquad
\mathcal C_M\mathcal P=+1~.
\eeq
This statement includes all $j=1$ spin-triplet sectors $L=0,1,2$ and does not require identifying a unique ${}^3P_1$ continuation at the photon pole.

By contrast, the scalar electric $j=1$ state has $L=1$ and $\mathcal C_M\mathcal P=-1$. The minimal electric fermion current is the ${}^3S_1$--${}^3D_1$ combination in Eq.~(\ref{eq: exact vector current LS}) and has parity $-1$; because $\mathcal C_M$ acts trivially on the electric pair, its combined eigenvalue is also $-1$. Thus each allowed electric initial state and the minimal magnetic spin-triplet final state lie in different eigenspaces of the exact symmetry.

Suppose, as in the reductio of the Introduction, that the amplitudes $\mathcal A_{s\to f}^{\min}$ and $\mathcal A_{f\to f}^{\min}$ exist with the specified minimal vertices as their photon-pole factorization data. The symmetry analysis then constrains the corresponding residues of these same hypothetical amplitudes to satisfy
\beq\label{eq: exact fermion residue zero}
R_{s\to f}\equiv
\underset{s=0}{\operatorname{Res}}\,\mathcal A_{s\to f}^{\min}=0~,
\qquad
R_{f\to f}\equiv
\underset{s=0}{\operatorname{Res}}\,\mathcal A_{f\to f}^{\min}=0~.
\eeq
These are the same pole coefficients that ordinary factorization evaluates in Section~\ref{sec: on-shell}. Under the existence assumption, each must equal the corresponding helicity sum $R_{\rm req}$. Since those sums are generically nonzero, we cannot retain an amplitude with a vanishing residue: such an object would fail to factorize into the specified nonzero minimal vertices. Instead, no ordinary local production amplitude with those factorization data exists. Independent regular four-point interactions may occupy symmetry-allowed entries of the complete basis, but Eq.~(\ref{eq: Laurent symmetry action}) shows that they do not change this pole constraint.

\subsection{Nonminimal Pauli three-point data}\label{sec: nonminimal Pauli data}

We now show explicitly that the fermionic obstruction is tied to the minimal magnetic three-point data. A regular four-point interaction cannot change the forbidden pole coefficient, but an independent magnetic form factor changes the factorization data themselves.

Besides the minimal spin-triplet amplitude, allow a nonzero Pauli form factor $\mathcal F_2(0)$. After translating the dual-photon  to the ordinary photon, an 
on-shell covariant amplitude is
\beq\label{eq: magnetic Pauli three point}
\mathcal M^{M,{\rm P}}_{3,h}
=c_P\,
\overline u(p_3)i\sigma^{\mu\nu}k_\nu\gamma^5v(p_4)\,
\epsilon_\mu^h(k)~,
\qquad k^2=0~,
\qquad [c_P]=-1~.
\eeq
Here $c_P$ is proportional to $\mathcal F_2(0)/m_M$, with the coupling normalization absorbed into its definition. Equation~(\ref{eq: magnetic Pauli three point}) specifies only on-shell three-point data; it does not assume an off-shell interaction or a single-potential description.

The helicity content follows directly from dimensional analysis. In the massless limit, the independent same-helicity amplitudes have the form
\beq\label{eq: magnetic Pauli helicity amplitudes}
\mathcal M^{M,{\rm P}}_3(3_M^-,4_{\overline M}^-,k^-)
\propto c_P\langle3k\rangle\langle4k\rangle~,
\qquad
\mathcal M^{M,{\rm P}}_3(3_M^+,4_{\overline M}^+,k^+)
\propto c_P[3k][4k]~,
\eeq
where the overall phases depend on the external-state convention. Unlike the minimal structures in Eq.~(\ref{eq: massless minimal photon amplitudes}), the same-helicity spinor polynomial has mass dimension two, so $c_P$ must carry inverse mass dimension. These amplitudes are therefore independent of the opposite-helicity spin-triplet amplitudes fixed by minimal coupling.

We can identify the exact finite-mass spin sector without choosing an off-shell completion. For equal monopole and antimonopole masses, the Dirac equations imply the pair-production Gordon identity. Using $\sigma^{\mu\nu}=i[\gamma^\mu,\gamma^\nu]/2$, $\bar u\slashed p_3=m_M\bar u$, $\slashed p_4v=-m_Mv$, and $\{\gamma^5,\gamma^\mu\}=0$, we first obtain
\beq
\begin{aligned}
\bar u\gamma^\mu\slashed k\gamma^5v
&=2p_3^\mu\bar u\gamma^5v~,\\
\bar u\slashed k\gamma^\mu\gamma^5v
&=2p_4^\mu\bar u\gamma^5v~.
\end{aligned}
\eeq
The mass terms cancel separately in the two reductions. Therefore
\beq\label{eq: Pauli pair Gordon identity}
\begin{aligned}
\overline u(p_3)i\sigma^{\mu\nu}k_\nu\gamma^5v(p_4)
&=-\frac{1}{2}\overline u(p_3)
\big(\gamma^\mu\slashed{k}-\slashed{k}\gamma^\mu\big)\gamma^5v(p_4)\\
&=-(p_3-p_4)^\mu\overline u(p_3)\gamma^5v(p_4)~.
\end{aligned}
\eeq
The fermion bilinear is a spin singlet, while the relative momentum supplies $L=1$. Thus the Pauli amplitude lies in the ${}^1P_1$ sector.

For electrically charged scalars in the initial state, define
\beq
Q^\mu\equiv(p_1-p_2)^\mu~,
\qquad
k^\mu=p_1^\mu+p_2^\mu=p_3^\mu+p_4^\mu~.
\eeq
Gluing the scalar electric three-point amplitude to Eq.~(\ref{eq: magnetic Pauli three point}) gives the photon-pole contribution, up to an overall phase convention,
\beq\label{eq: scalar to magnetic Pauli pole term}
\left.\mathcal A^{\rm P}_{s\rightarrow f}\right|_{\rm pole}
=\frac{e c_P}{s}\,Q_\mu\,
\overline u(p_3)i\sigma^{\mu\nu}k_\nu\gamma^5v(p_4)~.
\eeq
Using Eq.~(\ref{eq: Pauli pair Gordon identity}), we find
\beq\label{eq: reduced magnetic Pauli pole term}
\left.\mathcal A^{\rm P}_{s\rightarrow f}\right|_{\rm pole}
=-\frac{e c_P}{s}\,
\big[(p_1-p_2)\cdot(p_3-p_4)\big]\,
\overline u(p_3)\gamma^5v(p_4)~.
\eeq
In the high-energy limit, the two same-helicity components are therefore
\beq\label{eq: magnetic Pauli two structures}
\begin{aligned}
\mathcal A^{\rm P}_{--}
&\propto \frac{e c_P}{s}
\big[(p_1-p_2)\cdot(p_3-p_4)\big]\langle34\rangle~,\\
\mathcal A^{\rm P}_{++}
&\propto \frac{e c_P}{s}
\big[(p_1-p_2)\cdot(p_3-p_4)\big][34]~.
\end{aligned}
\eeq
The final-state factor $q_{\rm out}^\mu\overline u\gamma^5v$ is the spin-singlet $\widehat T_{10}={}^1P_1$ direction of the exact basis. It has $\mathcal C_M\mathcal P=-1$ and is the symmetry-allowed spin-singlet direction, so it matches the electric initial-state symmetry assignment. For generic kinematics, Eq.~(\ref{eq: reduced magnetic Pauli pole term}) is nonzero.

Thus the obstruction does not apply to arbitrary magnetic three-point data. It applies when the magnetic vertex is restricted to the minimal coupling spin-triplet amplitude. A nonzero Pauli form factor supplies independent spin-singlet factorization data and therefore defines a different production problem. It does not provide a regular completion of the residue generated by the minimal vertices.

The result can now be summarized directly. Discrete symmetry excludes the complete scalar $j=1$ basis without a minimal-coupling assumption. For fermions, the exact minimal magnetic vertex remains in the spin-triplet sector, whose $\mathcal C_M\mathcal P$ eigenvalue is opposite to that of either allowed electric initial state. Section~\ref{sec: nonminimal Pauli data} shows explicitly that an independent Pauli form factor supplies three-point data in the symmetry-allowed $\widehat T_{10}={}^1P_1$ sector. The complete basis sharpens the known vanishing result by showing that regular four-point additions cannot complete the tree-level construction, while nonminimal three-point data define a different factorization problem. The same minimal vertices occur in the known nonzero pairwise-covariant electric--magnetic scattering amplitudes; in the next section we evaluate those common three-point amplitudes directly on the production factorization locus, without crossing the four-point scattering amplitude.

\section{Production-channel factorization obstruction}\label{sec: on-shell}

Section~\ref{sec: CP} determined the symmetry-allowed ordinary $j=1$ production space. We now ask whether the minimal electric and magnetic three-point vertices can supply the photon-pole coefficient of an amplitude in that space. This is a conditional factorization test. We assume that the ordinary local tree-level amplitude $\mathcal A_{\min}$ exists with the stated three-point data and then compare the residue required by factorization with the residue allowed by the discrete symmetries. The same three-point vertices also occur in pairwise-covariant electric--magnetic scattering, but we will not continue or cross the four-point scattering amplitude. We evaluate the local three-point amplitudes directly on the complex production factorization locus.

We first distinguish the reference spinors from the reference vector. Let $\ket{\xi}$ and $|\xi]$ be a spinor pair, and define the corresponding null vector $\xi^\mu$ by
\beq\label{eq: reference vector definition}
\xi^\mu\equiv \bra{\xi}\sigma^\mu|\xi]~.
\eeq
The angle spinor $\ket{\xi}$ enters the three-point $x$-factor, while the reference vector $\xi^\mu$ enters the covariant off-shell representative below.

Let $k^\mu=-(p_1+p_2)^\mu$ be the photon momentum in a three-point amplitude containing a massive particle and antiparticle of mass $m$. For an admissible reference spinor $\ket{\xi}$ satisfying $\braket{\xi k}\neq0$, we define
\beq\label{eq:xfactor}
x_{12}\equiv x_{12}(\xi)
=\frac{\bra{\xi}(p_1-p_2)|k]}
{2m\braket{\xi k}}~.
\eeq
The $x$-factor is independent of the choice of $\ket{\xi}$ on three-point kinematics, as reviewed in Appendix~\ref{app: all e construct}. We use the same reference patch for the internal photon on the two sides of the factorization channel and abbreviate $x_{12}(\xi)$ and $x_{34}(\xi)$ by $x_{12}$ and $x_{34}$. The three-point amplitudes are then
\beq
 \includegraphics[height=2.4cm,valign=c]{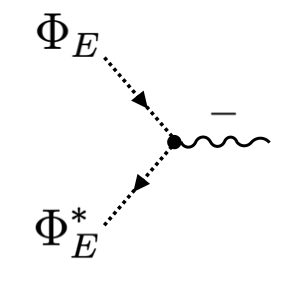}=c_{-}^E\frac{m}{x_{12}}~,
 \qquad
 \includegraphics[height=2.4cm,valign=c]{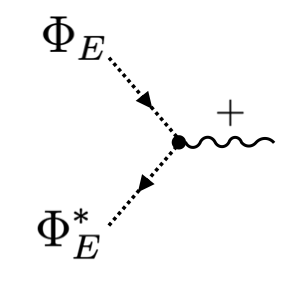}=c_{+}^E m x_{12}~,
\eeq
\beq
 \includegraphics[height=2.4cm,valign=c]{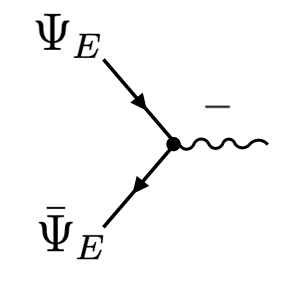}=c_{-}^E\frac{[\mathbf{12}]}{x_{12}}~,
 \qquad
 \includegraphics[height=2.4cm,valign=c]{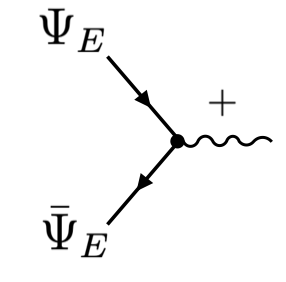}=c_{+}^E x_{12}\braket{\mathbf{12}}~,
\eeq
\beq
 \includegraphics[height=2.4cm,valign=c]{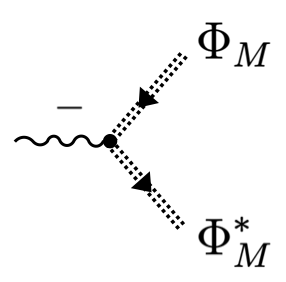}=c_{-}^M\frac{m'}{x_{34}}~,
 \qquad
 \includegraphics[height=2.4cm,valign=c]{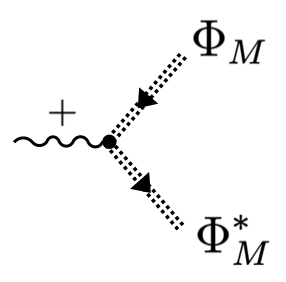}=c_{+}^M m' x_{34}~,
\eeq
\beq
 \includegraphics[height=2.4cm,valign=c]{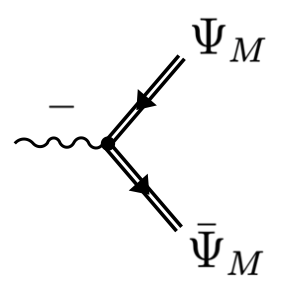}=c_{-}^M\frac{[\mathbf{34}]}{x_{34}}~,
 \qquad
 \includegraphics[height=2.4cm,valign=c]{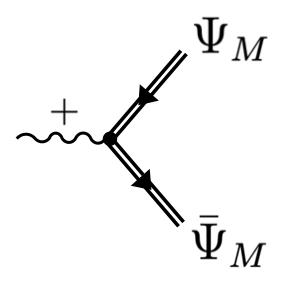}=c_{+}^M x_{34}\braket{\mathbf{34}}~.
\eeq
Here $c_\pm^E$ and $c_\pm^M$ are the reduced helicity coefficients defined in Eq.~(\ref{eq: magnetic interaction}), and all photon little-group weight is carried by the displayed $x$-factors. We denote the electric scalar and fermion by $\Phi_E$ and $\Psi_E$, and the magnetic scalar and fermion by $\Phi_M$ and $\Psi_M$. The corresponding all-electric or all-magnetic gluing constructions are reviewed in Appendix~\ref{app: all e construct}.

The magnitudes of the reduced coefficients do not affect the obstruction, so we set $|c_\pm^E|=|c_\pm^M|=1$. We choose
\beq
c_\pm^E=1~,
\qquad
c_\pm^M=\mp1~.
\eeq
Changing the common sign of the magnetic coefficients does not change the conclusion. With this convention,
\beq
 \includegraphics[height=2.4cm,valign=c]{escalarL.png}=\frac{m}{x_{12}}~,
 \qquad
 \includegraphics[height=2.5cm,valign=c]{escalarR.png}=m x_{12}~,
\eeq
\beq
 \includegraphics[height=2.4cm,valign=c]{efermiL.png}=\frac{[\mathbf{12}]}{x_{12}}~,
 \qquad
 \includegraphics[height=2.5cm,valign=c]{efermiR.png}=x_{12}\braket{\mathbf{12}}~,
\eeq
\beq
 \includegraphics[height=2.4cm,valign=c]{mscalarL.png}=\frac{m'}{x_{34}}~,
 \qquad
 \includegraphics[height=2.4cm,valign=c]{mscalarR.png}=-m'x_{34}~,
\eeq
\beq\label{eq: magnetic 3-point}
 \includegraphics[height=2.4cm,valign=c]{mfermiL.png}=\frac{[\mathbf{34}]}{x_{34}}~,
 \qquad
 \includegraphics[height=2.4cm,valign=c]{mfermiR.png}=-x_{34}\braket{\mathbf{34}}~.
\eeq
The relative minus sign for a positive-helicity photon is the electric--magnetic phase in Eq.~(\ref{eq: magnetic interaction}). It cancels when both vertices are magnetic, as required by duality, but it remains when an electric vertex is glued to a magnetic vertex. This relative phase is what enters the production residue.

\paragraph{Production-channel factorization.}
Return to the hypothetical amplitude $\mathcal A_{\min}$ introduced in Eq.~(\ref{eq: hypothetical minimal residue}). Since the minimal electric and magnetic vertices are assumed to be its factorization data, ordinary factorization on the complex divisor $k^2=0$ requires
\beq\label{eq: required production residue}
R\equiv
\underset{k^2=0}{\operatorname{Res}}\,\mathcal A_{\min}
=\sum_{h=\pm1}\mathcal M_h^E(1,2,k)\,
\mathcal M_{-h}^M(3,4,-k)
\equiv R_{\rm req}~,
\eeq
up to the conventional overall propagator normalization. The first equality identifies $R$ with the Laurent coefficient constrained in Section~\ref{sec: CP}. The second evaluates that same coefficient using the assumed three-point factorization data. Thus $R_{\rm req}$ is not an independently postulated four-point object; it is the value that $R$ must take if $\mathcal A_{\min}$ exists.

The divisor $k^2=0$ lies outside the real massive production region. This does not assert that the full confined process is physically described by the hard approximation at $k^2=0$. Rather, it tests the assumption that the hard production kernel admits an ordinary rational local completion. Under that assumption, the complex Laurent coefficient is well defined and must factorize as in Eq.~(\ref{eq: required production residue}).

For scalar production, the helicity sum is
\beq\label{eq: gluing production}
R_{\rm req}=
\includegraphics[height=2.4cm,valign=c]{escalarL.png}
\includegraphics[height=2.4cm,valign=c]{mscalarR.png}
+
\includegraphics[height=2.4cm,valign=c]{escalarR.png}
\includegraphics[height=2.4cm,valign=c]{mscalarL.png}~.
\eeq
The internal momenta are outgoing as $k$ on one subamplitude and $-k$ on the other. Since helicity is the spin projection along the momentum, the same physical photon state is labeled by $h$ and $-h$ on the two sides. We therefore obtain
\begin{itemize}
    \item scalar to scalar,
    \beq\label{eq: scalar xx}
    R_{s\to s}=mm'\left(\frac{x_{34}}{x_{12}}-\frac{x_{12}}{x_{34}}\right)~;
    \eeq

    \item scalar to fermion,
    \beq
    R_{s\to f}=m\left(
    \frac{x_{34}}{x_{12}}\braket{\mathbf{34}}
    -\frac{x_{12}}{x_{34}}[\mathbf{34}]
    \right)~;
    \eeq

    \item fermion to fermion,
    \beq
    R_{f\to f}=
    \frac{x_{34}}{x_{12}}[\mathbf{12}]\braket{\mathbf{34}}
    -\frac{x_{12}}{x_{34}}\braket{\mathbf{12}}[\mathbf{34}]~.
    \eeq
\end{itemize}
These expressions are obtained by evaluating the local three-point amplitudes on complex on-shell three-point kinematics and gluing them across the two physical photon helicities. No four-point scattering amplitude has been crossed or analytically continued.

\paragraph{Reference independence of the pole coefficient.}
We now verify that the helicity sum defines a physical Laurent coefficient. For the scalar and fermion channels, define the conserved electric and magnetic currents
\beq\label{eq: J K definition}
\begin{aligned}
    J_s^\mu&\equiv \frac{(p_1-p_2)^\mu}{\sqrt{2}}~,
    &K_s^\mu&\equiv \frac{(p_3-p_4)^\mu}{\sqrt{2}}~,\\
    J_f^\mu&\equiv
    \frac{\bra{\mathbf{1}}\sigma^\mu|\mathbf{2}]
    +\bra{\mathbf{2}}\sigma^\mu|\mathbf{1}]}{\sqrt{2}}~,
    &K_f^\mu&\equiv
    \frac{\bra{\mathbf{3}}\sigma^\mu|\mathbf{4}]
    +\bra{\mathbf{4}}\sigma^\mu|\mathbf{3}]}{\sqrt{2}}~.
\end{aligned}
\eeq
For each channel we choose the corresponding pair $(J,K)$ from Eq.~(\ref{eq: J K definition}). These currents obey $k\cdot J=k\cdot K=0$.

An off-shell representative of the helicity sum can be written in terms of the reference vector $\xi^\mu$ as
\beq\label{eq: R = epsilon}
N_\xi\equiv
\frac{-i\varepsilon_{\mu\nu\alpha\beta}
 k^\mu \xi^\nu J^\alpha K^\beta}
{k\cdot \xi}~.
\eeq
For an admissible reference patch, $k\cdot \xi\neq0$. On the factorization locus, $N_\xi$ equals the helicity sum in Eq.~(\ref{eq: required production residue}), up to the common normalization already suppressed there.

The off-shell representative depends on the reference spinors, but its value on the photon pole does not. Let $\ket{\chi}$ and $|\chi]$ define a second null reference vector $\chi^\mu$. The Schouten identity derived in Appendix~\ref{app: reference continuation} gives the exact linear relation
\beq\label{eq: reference ambiguity}
N_\xi-N_\chi
=i k^2
\frac{\varepsilon_{\mu\nu\alpha\beta}
 \xi^\mu \chi^\nu J^\alpha K^\beta}
{(k\cdot \xi)(k\cdot \chi)}~.
\eeq
On the overlap of two admissible reference patches, both denominators remain finite as $k^2\to0$. Therefore
\beq
N_\xi\big|_{k^2=0}=N_\chi\big|_{k^2=0}=R_{\rm req}~.
\eeq
Thus the Laurent coefficient is reference independent even though a chosen off-shell representative is not. If $k\cdot \xi$ vanishes at a special point of the factorization divisor, the spinor pair $(\ket{\xi},|\xi])$ no longer defines an admissible polarization patch, so we extract the residue in another patch. A correlated limit in which $k\cdot \xi$ and $k^2$ vanish together probes a singular representative rather than an ambiguity in the residue.

The reference spinors $\ket{\xi}$ and $|\xi]$, together with the reference vector $\xi^\mu$ constructed from them, specify the photon-polarization gauge choice. They are not recursion-shift data. A change of recursion shift may redistribute spurious denominators between pole terms and a boundary term, but it cannot change the reference-independent Laurent coefficient of the final amplitude.

We can now test whether this coefficient vanishes generically. Squaring Eq.~(\ref{eq: R = epsilon}) on the factorization locus and using $k^2=k\cdot J=k\cdot K=0$ gives
\beq\label{eq: invariant residue norm}
R_{\rm req}^{\,2}=(J\cdot K)^2-J^2K^2~,
\eeq
up to the common coupling normalization. Since $J$ and $K$ are defined modulo shifts by the null gauge direction,
\beq
J\sim J+\alpha k~,
\qquad
K\sim K+\beta k~,
\eeq
the right-hand side is the Gram determinant of their classes in the physical transverse space $k^\perp/\operatorname{span}\{k\}$. It is nonzero unless the two classes are proportional. Thus $R_{\rm req}$ is generically nonzero and vanishes only on a lower-dimensional kinematic locus.

\paragraph{No regular local completion.}
At fixed order in a local derivative expansion, every $j=1$ Laurent coefficient is a polynomial covariant in the external momenta and wavefunctions, modulo the external equations of motion. We can therefore expand an ordinary local production amplitude in the complete basis $B_i$ of Table~\ref{tb: massive Mj} as
\beq\label{eq: physical completion ansatz}
\mathcal A^{(j=1)}
=\sum_i\left(\frac{r_i}{k^2}+f_i(k^2)\right)B_i~,
\eeq
where each $f_i$ is polynomial at fixed order in the derivative expansion. Factorization fixes a generically nonzero set of coefficients $r_i$ through Eq.~(\ref{eq: required production residue}). Section~\ref{sec: CP}, however, shows that the discrete symmetries require the pole coefficient of the same hypothetical amplitude to lie in the vanishing production subspace. Since Eq.~(\ref{eq: Laurent symmetry action}) constrains the pole and regular Laurent coefficients separately, no choice of the regular functions $f_i$ can repair the mismatch. Thus the specified minimal vertices cannot be assembled into an ordinary local tree-level production amplitude under the assumptions stated in the Introduction.

The result can be avoided only by changing one of those assumptions. A nonminimal Pauli form factor changes the three-point data. Additional light states add poles or cuts, while nonlocal string or confining dynamics can invalidate an isolated local hard amplitude. Heavy states whose low-energy effect is only a regular local four-point interaction are already included in the functions $f_i$ and therefore do not help. A different asymptotic-state representation changes the symmetry classification.

Electric--magnetic scattering realizes the last possibility. Its asymptotic states carry pairwise little-group weight, so the known nonzero scattering amplitude belongs to a different four-point representation. It is not a crossed realization of the forbidden ordinary production residue. Pairwise covariance is not needed to prove the production obstruction, but it explains why that obstruction is not an inconsistency of the individual three-point vertices.

For completeness, Appendix~\ref{app: MM check} asks whether the scalar factor introduced in Ref.~\cite{Moynihan:2020gxj} for electric--magnetic scattering could instead be used as a production-channel continuation. Reference~\cite{Moynihan:2020gxj} did not make this claim; the appendix records only the result of that conditional test. A pairwise-covariant scattering amplitude can also be organized in the partial-wave basis of Refs.~\cite{Csaki:2020inw,Csaki:2020yei}. For $t$-channel scattering, however, the individual terms in such an expansion are not residues of physical exchanged particles and should not be interpreted as ordinary on-shell factorization channels.


\section{Conclusion}\label{sec: conclusion}

We have shown that the tree-level single-photon amplitude for electric-pair-to-magnetic-pair production cannot be constructed from the minimal electric and magnetic three-point vertices under the assumptions stated in the Introduction. The complete ordinary $j=1$ production basis contains one scalar-to-scalar structure, four scalar-to-fermion structures, and sixteen fermion-to-fermion structures. Assuming that the amplitude exists, the discrete symmetries require its photon-pole coefficient $R$ to vanish, while factorization of that same amplitude into the specified minimal vertices requires $R=R_{\rm req}\neq0$. Since the symmetry leaves the channel invariant, Eqs.~(\ref{eq: Laurent symmetry separation}) and~(\ref{eq: Laurent symmetry action}) constrain the pole and regular terms separately. Thus no finite-derivative regular local four-point interaction can repair the contradiction. This is an obstruction to the existence of the ordinary local tree-level amplitude, rather than a failure of a particular recursion relation.

For fermions, the exact finite-mass vector current is a fixed ${}^3S_1$--${}^3D_1$ combination. Minimal electric--magnetic duality changes the photon-helicity phase but leaves the massive spin indices unchanged, so the magnetic pair remains spin triplet. Every magnetic spin-triplet state has $\mathcal C_M\mathcal P=+1$, while the electric initial state has eigenvalue $-1$. An independent Pauli form factor avoids this obstruction by supplying three-point data in the symmetry-allowed spin-singlet sector. Whether this form factor is nonzero depends on the microscopic theory.

The same minimal vertices occur in known nonzero electric--magnetic scattering amplitudes because scattering states carry pairwise little-group weight, while production states transform in the ordinary tensor product of one-particle representations. The production result is therefore channel dependent and does not imply an inconsistency of the individual three-point vertices. 

A nonzero one-photon production kernel must therefore change the three-point data, add light singularities, involve nonlocal confining dynamics, or use a different asymptotic-state representation. Multiphoton channels are not subject to this constraint. In particular, the tree-level two-photon amplitude is nonzero, and its leading high-energy term has $j=0$~\cite{Terning:2020dzg}.

\section*{Acknowledgments}

We thank Chris Verhaaren, Joshua Newey, Tim Cohen, Markus Luty, and Hsin-Chia Cheng for insightful discussions.


\appendix
\numberwithin{equation}{section}
\renewcommand{\theequation}{\thesection.\arabic{equation}}

\section{Useful properties of spinor-helicity variables}\label{app: formula}
\beq
\Tr{(\sigma^\mu\bar{\sigma}^\nu)}=2\eta^{\mu\nu}~,\quad\varepsilon_{\alpha\beta}\varepsilon^{\beta\gamma}=\delta_\alpha^\gamma~.
\eeq
\beq
\sigma^\mu_{\alpha\dot{\alpha}}\sigma_{\mu\beta\dot{\beta}}=2\varepsilon_{\alpha\beta}\varepsilon_{\dot{\alpha}\dot{\beta}}~,\quad\sigma^\mu_{\alpha\dot{\alpha}}\bar{\sigma}_{\mu}^{\dot{\beta}\beta}=2\delta^\beta_\alpha\delta^{\dot{\beta}}_{\dot{\alpha}}~.
\eeq
\beq
\left(\sigma^\mu\bar{\sigma}^\nu+\sigma^\nu\bar{\sigma}^\mu\right)^\beta_\alpha=2\eta^{\mu\nu}\delta^\beta_\alpha~.
\eeq
\beq\label{eq: schouten}
\begin{aligned}
    \text{Schouten identity:}\quad\quad\quad\varepsilon_{\alpha\beta}\varepsilon_{\gamma\delta}=&\varepsilon_{\alpha\gamma}\varepsilon_{\beta\delta}-\varepsilon_{\alpha\delta}\varepsilon_{\beta\gamma}\\
    \Rightarrow |p][qk]+|q][kp]+|k][pq]&=0~,\quad\ket{p}\braket{qk}+\ket{q}\braket{kp}+\ket{k}\braket{pq}=0~.
\end{aligned}
\eeq
\beq
p_i\ket{i^I}=m_i|i^I]~,\quad p_i|i^I]=m_i\ket{i^I}~,\quad \bra{i^I}p_i=-m_i[i^I|~,\quad [i^I|p_i=-m_i\bra{i^I}~.
\eeq
\beq
|i^I][i_I|=m_i\mathds{1}~,\quad\ket{i_I}\bra{i^I}=m_i\mathds{1}~.
\eeq
\beq\label{eq: trace}
\begin{aligned}
&\Tr{(\sigma_\mu\bar{\sigma}_\nu\sigma_\alpha\bar{\sigma}_\beta)}=2(\eta_{\mu\nu}\eta_{\alpha\beta}-\eta_{\mu\alpha}\eta_{\nu\beta}+\eta_{\mu\beta}\eta_{\nu\alpha}+i\varepsilon_{\mu\nu\alpha\beta})~,\\
&\Tr{(\bar{\sigma}_\mu\sigma_\nu\bar{\sigma}_\alpha\sigma_\beta)}=2(\eta_{\mu\nu}\eta_{\alpha\beta}-\eta_{\mu\alpha}\eta_{\nu\beta}+\eta_{\mu\beta}\eta_{\nu\alpha}-i\varepsilon_{\mu\nu\alpha\beta})~.
\end{aligned}
\eeq
\beq
\bra{i}p|j]=[j|p\ket{i}~,\quad\braket{qk}[kq]=2q\cdot k~,\quad\varepsilon_{\dot{\alpha}\dot{\beta}}\left(p\ket{q}\right)^{\dot{\beta}}=-(\bra{q}p)_{\dot{\alpha}}~.
\eeq
\beq
[j^Ij^J]=m_j\varepsilon^{IJ}~,\quad\braket{j_Ij_J}=m_j\varepsilon_{IJ}~.
\eeq
\beq
\varepsilon^{\alpha\beta}p^{\dot{\alpha}\gamma}=\varepsilon^{\alpha\gamma}p^{\dot{\alpha}\beta}+\varepsilon^{\gamma\beta}p^{\dot{\alpha}\alpha}~,\quad\varepsilon^{\dot{\alpha}\dot{\beta}}p^{\dot{\gamma}\alpha}=\varepsilon^{\dot{\alpha}\dot{\gamma}}p^{\dot{\beta}\alpha}+\varepsilon^{\dot{\gamma}\dot{\beta}}p^{\dot{\alpha}\alpha}~.
\eeq
\beq
p^{\dot{\alpha}\beta}\bra{q}^\alpha =(p\ket{q})^{\dot{\alpha}}\varepsilon^{\alpha\beta}+p^{\dot{\alpha}\alpha}\bra{q}^\beta~,\quad p^{\dot{\beta}\alpha}|q]^{\dot{\alpha}} =([q|p)^{\alpha}\varepsilon^{\dot{\alpha}\dot{\beta}}+p^{\dot{\alpha}\alpha}|q]^{\dot{\beta}}~.
\eeq
\beq
p^{\dot{\alpha}\alpha}q^{\dot{\beta}\beta}=p^{\dot{\beta}\alpha}q^{\dot{\alpha}\beta}+\varepsilon^{\dot{\alpha}\dot{\beta}}(pq)^\alpha_\sigma\varepsilon^{\sigma\beta}=p^{\dot{\alpha}\beta}q^{\dot{\beta}\alpha}-\varepsilon^{\alpha\beta}(qp)^{\dot{\beta}}{}_{\dot{\sigma}}\varepsilon^{\dot{\sigma}\dot{\alpha}}~.
\eeq
\beq
p^{\dot{\alpha}\alpha}\varepsilon_{\alpha\beta}q^{\dot{\beta}\beta}=(qp)^{\dot{\beta}}_{\dot{\gamma}}\varepsilon^{\dot{\gamma}\dot{\alpha}}~,\quad p^{\dot{\alpha}\alpha}\varepsilon_{\dot{\alpha}\dot{\beta}}q^{\dot{\beta}\beta}=(pq)^{\beta}_{\gamma}\varepsilon^{\gamma\alpha}~.
\eeq
\beq
p_{\rho\dot{\alpha}}\varepsilon^{\rho\lambda}q_{\lambda\dot{\beta}}=(qp)^{\dot{\rho}}_{\dot{\alpha}}\varepsilon_{\dot{\rho}\dot{\beta}}~,\quad p_{\alpha\dot{\rho}}\varepsilon^{\dot{\rho}\dot{\lambda}}q_{\beta\dot{\lambda}}=-(pq)^\rho_\alpha\varepsilon_{\rho\beta}~.
\eeq
\beq
(pq)^{\dot{\alpha}}_{\dot{\beta}}\varepsilon^{\rho\sigma}=(\varepsilon^{\rho\lambda}p^{\dot{\alpha}\sigma}+\varepsilon^{\lambda\sigma}p^{\dot{\alpha}\rho})q_{\lambda\dot{\beta}}~,\quad(pq)^{\alpha}_{\beta}\varepsilon^{\dot{\rho}\dot{\sigma}}=(\varepsilon^{\dot{\rho}\dot{\lambda}}p^{\dot{\sigma}\alpha}+\varepsilon^{\dot{\lambda}\dot{\sigma}}p^{\dot{\rho}\alpha})q_{\beta\dot{\lambda}}~.
\eeq
\beq
\varepsilon^{\rho\alpha}\varepsilon_{\lambda\beta}(pq)^\beta_\alpha=-(qp)^\rho_\lambda~,\quad\varepsilon^{\dot{\rho}\dot{\alpha}}\varepsilon_{\dot{\lambda}\dot{\beta}}(pq)^{\dot{\beta}}_{\dot{\alpha}}=-(qp)^{\dot{\rho}}_{\dot{\lambda}}~.
\eeq

\section{Parity transformation of massive spinors}\label{app: parity}

To constrain an amplitude by parity conservation, we examine the parity transformation of the $S$-matrix. Let $p_j^\mu=(E_j,\V{p}_j)$ and define its parity transform by $\bar{p}_j^\mu=(E_j,-\V{p}_j)$. If parity is conserved, the $S$-matrix elements will satisfy \cite{Weinberg:1995mt}:
\beq\label{eq: S matrix parity}
\bra{p_1',p_2',\ldots}\hat{S}\ket{p_1,p_2,\ldots}=\eta_{1'}^*\eta_{2'}^*\cdots\eta_1\eta_2 \bra{\bar{p}_1',\bar{p}_2',\ldots}\hat{S}\ket{\bar{p}_1,\bar{p}_2,\ldots}~,
\eeq
where $p_i'$ are the outgoing momenta, $p_i$ are the incoming momenta, and $\eta_i$ are phases. We suppress the spin labels, but the spin polarization of each particle is the same on both sides of Eq.~(\ref{eq: S matrix parity}), while their helicities change sign. Equation~(\ref{eq: S matrix parity}) relates the $S$-matrix evaluated with incoming momenta $p_i$ to that evaluated with their parity transforms $\bar p_i$. Since the amplitude will be expressed with spinor-helicity variables, we need to know how to express them with a parity-conjugated momentum $\bar{p}$. In other words, we need to know how $\braket{\bar{p}\bar{q}}$, $[\bar{p}\bar{q}]$ are related to $\braket{pq}$, $[pq]$ to discuss parity using spinor-helicity techniques.

We begin with explicit formulas for the spinor-helicity variables \cite{Liu:2022alx}. Let $p=|\mathbf{p}|$ be the magnitude of the three-momentum and $(\theta,\phi)$ denote its direction in spherical coordinates. Defining $c\equiv e^{i\phi/2}\cos{(\theta/2)}$ and $s\equiv e^{i\phi/2}\sin{(\theta/2)}$, we have:
\beq
\lambda^I_\alpha = \begin{pmatrix}
    \sqrt{E-p} ~c^* &- \sqrt{E+p} ~s^*\\
    \sqrt{E-p}~ s & \sqrt{E+p} ~c
\end{pmatrix}~,\quad
\Tilde{\lambda}_{I\dot{\alpha}}=\begin{pmatrix}
    \sqrt{E-p} ~c & \sqrt{E-p} ~s^*\\
    -\sqrt{E+p}~ s & \sqrt{E+p} ~c^*
\end{pmatrix}~.
\eeq
For null momenta, these formulas reduce to
\beq
\lambda_\alpha=\sqrt{2E}\begin{pmatrix}
    -s^* \\
    c
\end{pmatrix}~,\quad
\Tilde{\lambda}_{\dot{\alpha}}=\sqrt{2E}\begin{pmatrix}
    -s & c^*
\end{pmatrix}~.
\eeq
Under parity, $p^\mu\rightarrow\bar{p}^\mu$ and $(\theta,\phi)\rightarrow(\pi-\theta,\phi+\pi)$. Substituting the transformed angular coordinates into the explicit spinor formulas, we find:
\beq\label{eq: contracted parity spinor}
\begin{aligned}
    \braket{\bar{p}\bar{q}}=[pq]~,&\quad[\bar{p}\bar{q}]=\braket{pq}\\
    \braket{\bar{p}\bar{q}^I}=[pq^{(-I)}]~,&\quad[\bar{p}\bar{q}^I]=\braket{pq^{(-I)}}\\
    \braket{\bar{p}^I\bar{q}^J}=[p^{(-I)}q^{(-J)}]~,&\quad[\bar{p}^I\bar{q}^J]=\braket{p^{(-I)}q^{(-J)}}~.
\end{aligned}
\eeq
Here, a minus sign in front of the $SU(2)$ index indicates the exchange $I=1\leftrightarrow2$, and hence a helicity flip. Note that for expressions involving contracted $SU(2)$ indices we have
\beq
\begin{aligned}
    \bra{\bar{i}^{I}}\bar{p}|\bar{j}^J]=&\braket{\bar{i}^I\bar{p}^K}[\bar{p}_K\bar{j}^J]\\
    =&[i^{(-I)}p^{(-K)}]\braket{p_{(-K)}j^{(-J)}}\\
    =&[i^{(-I)}p^{K}]\braket{p_{K}j^{(-J)}}\\
    =&-[i^{(-I)}p_{K}]\braket{p^{K}j^{(-~J)}}\\
    =&-[i^{(-I)}|p\ket{j^{(-J)}}~,
\end{aligned}
\eeq
where $|p^K]\bra{p_K}=\varepsilon^{KL}|p_L]\bra{p_K}=-\varepsilon^{LK}|p_L]\bra{p_K}=-|p_L]\bra{p^L}$ because $\varepsilon_{LK}$ is antisymmetric.

In the bold notation, we display the helicity flip.
\beq
\braket{\bar{i\vphantom{\mathbf{j}}}\,\bar{\mathbf{j}}} \Big| _{\{h\}}=-[i\,\mathbf{j}]\Big|_{\{-h\}}~,\quad\text{etc.}
\eeq
In summary, parity exchanges $\braket{~}\leftrightarrow[~]$ while reversing all particle helicities; the pairwise helicity $h_{ij}$ is not included in $\{h\}$. Every pair of contracted $SU(2)$ indices gives a minus sign. Since amplitudes are always expressed with fully contracted spinors, Eq.~(\ref{eq: contracted parity spinor}) is all we need.



\section{Reference independence of the pole coefficient}\label{app: reference continuation}

This appendix proves Eq.~(\ref{eq: reference ambiguity}) directly. The reference spinor pair $\ket{\xi}$ and $|\xi]$ defines the null reference vector $\xi^\mu$ in Eq.~(\ref{eq: reference vector definition}). A second spinor pair $\ket{\chi}$ and $|\chi]$ similarly defines $\chi^\mu$. The spinors and vectors carry the same labels, but the ket or square-bracket notation distinguishes the spinors from the Lorentz-indexed vectors.

For any vector $q^\mu$, define
\beq
E(q)\equiv
\varepsilon_{\mu\nu\alpha\beta}
 k^\mu q^\nu J^\alpha K^\beta~,
\qquad
N(q)\equiv-\frac{iE(q)}{k\cdot q}~.
\eeq
Thus $N_\xi=N(\xi)$ and $N_\chi=N(\chi)$. We now apply the five-vector Schouten identity to $k$, $\xi$, $\chi$, $J$, and $K$:
\beq
\begin{aligned}
0={}&k^\lambda\varepsilon(\xi,\chi,J,K)
-\xi^\lambda\varepsilon(k,\chi,J,K)
+\chi^\lambda\varepsilon(k,\xi,J,K)\\
&-J^\lambda\varepsilon(k,\xi,\chi,K)
+K^\lambda\varepsilon(k,\xi,\chi,J)~,
\end{aligned}
\eeq
where $\varepsilon(a,b,c,d)\equiv
\varepsilon_{\mu\nu\alpha\beta}a^\mu b^\nu c^\alpha d^\beta$. Contracting this identity with $k_\lambda$ and using $k\cdot J=k\cdot K=0$, we obtain
\beq\label{eq: Schouten reference identity}
(k\cdot \chi)E(\xi)
-(k\cdot \xi)E(\chi)
=-k^2\varepsilon(\xi,\chi,J,K)~.
\eeq
Dividing by $(k\cdot \xi)(k\cdot \chi)$ gives
\beq
N_\xi-N_\chi
=i k^2
\frac{\varepsilon_{\mu\nu\alpha\beta}
 \xi^\mu \chi^\nu J^\alpha K^\beta}
{(k\cdot \xi)(k\cdot \chi)}~,
\eeq
which is Eq.~(\ref{eq: reference ambiguity}). This is an exact linear identity; it does not require differentiating with respect to a reference spinor or squaring the resulting variation.

On the overlap of two admissible polarization patches, $k\cdot \xi$ and $k\cdot \chi$ are nonzero. The right-hand side therefore vanishes at $k^2=0$, so the pole coefficient is independent of the reference spinors. If either denominator vanishes at a special point, the corresponding spinor pair has left its admissible polarization patch, and we use another patch before extracting the residue. No statement about a local off-shell completion is needed.

\section{A conditional square-root check}\label{app: MM check}
Reference~\cite{Moynihan:2020gxj} introduced the scalar factor
\beq\label{eq: square root}
\sqrt{\left(\frac{t-u}{2}\right)^2-4m^2m'^2}
\eeq
for electric--magnetic scattering and did not propose it as a production amplitude. If one nevertheless asks whether this factor could be used as a continuation of the single-photon production residue, its angular dependence provides a direct check. It ceases to be a pure $j=1$ function away from the factorization locus. In the center-of-momentum frame, with $x=\cos\theta$,
\beq
t-u=4|\mathbf p_{\rm in}||\mathbf p_{\rm out}|x~.
\eeq
For $x\neq0$ and
\beq
\frac{m^2m'^2}{|\mathbf p_{\rm in}|^2|\mathbf p_{\rm out}|^2x^2}\ll1~,
\eeq
the high-energy expansion has the form
\beq\label{eq: sqrt angular expansion}
\sqrt{\left(\frac{t-u}{2}\right)^2-4m^2m'^2}
=2|\mathbf p_{\rm in}||\mathbf p_{\rm out}|x
-\frac{m^2m'^2}{|\mathbf p_{\rm in}||\mathbf p_{\rm out}|x}+\cdots~.
\eeq
The first term is proportional to $P_1(x)$ and is a $j=1$ partial wave. The Legendre equation for $\ell=1$ is
\begin{equation}
\left[(1-x^2)\frac{d^2}{dx^2}-2x\frac{d}{dx}+2\right]f(x)=0~.
\end{equation}
The first mass correction is proportional to $1/x$. Acting with this operator gives
\beq\label{eq: inverse x mismatch}
\left[(1-x^2)\frac{d^2}{dx^2}-2x\frac{d}{dx}+2\right]\frac{1}{x}
=\frac{2(1+x^2)}{x^3}\neq0~.
\eeq
Thus the $O(m^2m'^2)$ term is the first explicit obstruction. The function $1/x$ contains a mixture of partial waves rather than a pure higher-$j$ component, so Eq.~(\ref{eq: square root}) is not a $j=1$ continuation of the production residue away from the factorization locus. This production-channel statement is secondary to, and logically distinct from, the pairwise-little-group objection to using the same expression as a complete scattering amplitude.


\section{Exact $j=1$ relative-momentum decomposition}\label{app: details}

The purpose of this appendix is to isolate the angular-momentum components without using an approximate deformation of the external momenta. We keep
\beq
P^\mu\equiv(P^T)^\mu=p_1^\mu+p_2^\mu~,
\qquad
s=P^2~,
\qquad
q^\mu=\frac{1}{2}(p_1-p_2)^\mu~,
\eeq
so equal masses imply
\beq
P\cdot q=0~,
\qquad
q^2=m^2-\frac{s}{4}~.
\eeq
The tensor
\beq
\Pi^{\mu\nu}=\eta^{\mu\nu}-\frac{P^\mu P^\nu}{s}
\eeq
projects onto the three-dimensional space transverse to $P$. In the center-of-momentum frame, this is the ordinary spatial rotation space.

The exact radial deformation is
\beq\label{eq: exact appendix deformation}
q\longrightarrow q(r)=rq~,
\qquad
p_{1,2}(r)=\frac{1}{2}P\pm rq~,
\qquad
m_r^2=\frac{s}{4}+r^2\left(m^2-\frac{s}{4}\right)~.
\eeq
It obeys
\beq
p_1(r)+p_2(r)=P~,
\qquad
p_1(r)^2=p_2(r)^2=m_r^2
\eeq
for every complex $r$. Thus the deformation changes only the radial coordinate in the transverse momentum space. It does not change the direction of $q$, and it commutes with the angular action of the Pauli--Lubanski operator. For fermionic states we use this deformation only to grade the orbital tensors, so no choice of branch for $m_r$ or deformation of the external spinors is required.

\subsection{Particles 1 and 2 are scalar}\label{app: scalar detail}

A polynomial in $q$ can be decomposed into transverse symmetric-traceless tensors. We write
\beq\label{eq: harmonic decomposition}
q^{\mu_1}\cdots q^{\mu_N}
=\sum_{k=0}^{\lfloor N/2\rfloor}(q^2)^k
q^{\langle\mu_1}\cdots q^{\mu_{N-2k}\rangle}\,
\Pi^{(k)}~,
\eeq
where the brackets denote the symmetric-traceless projection in the space transverse to $P$, and $\Pi^{(k)}$ denotes the corresponding products of transverse metrics. The precise normalization of the trace terms is not needed. The rank-$\ell$ tensor
\beq
q^{\langle\mu_1}\cdots q^{\mu_\ell\rangle}
\eeq
is an irreducible spin-$\ell$ representation of the rotation group. Therefore,
\beq\label{eq: W2 harmonic}
W^2 q^{\langle\mu_1}\cdots q^{\mu_\ell\rangle}
=-s\ell(\ell+1)
q^{\langle\mu_1}\cdots q^{\mu_\ell\rangle}~.
\eeq
Under Eq.~(\ref{eq: exact appendix deformation}), a homogeneous term of degree $N$ scales as $r^N$. More importantly, tensors of different rank $\ell$ cannot cancel because they belong to inequivalent irreducible representations. This is the exact content that the earlier approximate scaling argument was intended to capture.

For scalar particles 1 and 2, the total angular momentum is purely orbital. The $j=1$ condition therefore keeps only $\ell=1$. Every such term has the form
\beq
(q^2)^k q^\mu F_{k\mu}(p_3,p_4)~.
\eeq
Since $q^2=m^2-s/4$ is a scalar fixed by the external invariants, its powers can be absorbed into the coefficient. Hence
\beq\label{eq: scalar exact j1}
\mathcal M^{(j=1)}=q^\mu F_\mu(p_3,p_4)
=\frac{1}{2}(p_1-p_2)^\mu F_\mu(p_3,p_4)~.
\eeq
Absorbing the factor of $1/2$ into $F_\mu$ gives Eq.~(\ref{eq: scalar final ansatz}). This argument also shows why higher powers of the relative momentum do not define additional $j=1$ structures: their rank-one trace components reduce to Eq.~(\ref{eq: scalar exact j1}), while their traceless components have $j\neq1$.

\subsection{Particles 1 and 2 are fermion}\label{app: fermion detail}

For two spin-$1/2$ particles, the intrinsic spin decomposes into a singlet $\Sigma$ and a triplet $S^\mu$, with
\beq
P\cdot S=0~.
\eeq
The orbital tensors are the same transverse harmonics used above. The addition rule
\beq
|L-S|\leq j\leq L+S
\eeq
shows that $j=1$ can occur only for
\beq
(L,S)=(1,0),\qquad(0,1),\qquad(1,1),\qquad(2,1)~.
\eeq
A convenient covariant vector for each sector is
\beq\label{eq: exact fermion covariants}
\begin{aligned}
T_{10}^\mu&=q^\mu\Sigma~,\\
T_{01}^\mu&=S^\mu~,\\
T_{11}^\mu&=\frac{1}{\sqrt{s}}\epsilon^{\mu\nu\rho\sigma}P_\nu q_\rho S_\sigma~,\\
T_{21}^\mu&=q^\mu(q\cdot S)-\frac{1}{3}q^2S^\mu~.
\end{aligned}
\eeq
The last line is obtained by coupling the rank-two transverse symmetric-traceless tensor
\beq
Q^{\mu\nu}=q^\mu q^\nu-\frac{1}{3}q^2\Pi^{\mu\nu}
\eeq
to the spin triplet. Each $T_{LS}^\mu$ transforms as $j=1$, and the four tensors exhaust the Clebsch--Gordan decomposition. Extra factors of $q^2$ again change only the scalar coefficient. Thus the most general initial-state dependence can be written as
\beq\label{eq: exact fermion j1}
\mathcal M^{(j=1)}
=\sum_{(L,S)}T_{LS}^\mu F^{LS}_\mu(p_3,p_4)~,
\qquad
(L,S)=(1,0),(0,1),(1,1),(2,1)~.
\eeq

We now relate Eq.~(\ref{eq: exact fermion j1}) to the massive spinor ansatz in Eq.~(\ref{eq: fermion ansatz}). The spin singlet is represented by the antisymmetric parts of the same-chirality products, proportional to $\braket{\mathbf{12}}$ or $[\mathbf{12}]$. The spin triplet is represented by their symmetric parts together with the mixed-chirality bilinears. We then replace $q=(p_1-p_2)/2$ and use the massive Dirac equations to move every occurrence of $p_1$ or $p_2$ next to an external spinor. Schouten identities remove the remaining redundant spinor contractions. In this reduction, the $L=2$ tensor in Eq.~(\ref{eq: exact fermion covariants}) is not an additional quadratic coefficient in Eq.~(\ref{eq: fermion ansatz}); the Dirac equations reduce it to the displayed $N=0$ and $N=1$ spinor structures.

The resulting spinor basis can be chosen to satisfy
\beq\label{eq: exact fermion coefficient reduction}
g^{N}_{\beta\gamma}=\bar g^{N}_{\dot\beta\dot\gamma}=0\quad(N\geq2)~,
\qquad
h^{N}_{\beta\dot\gamma}=\bar h^{N}_{\dot\beta\gamma}=0\quad(N\neq0)~.
\eeq
For the remaining same-chirality terms, the antisymmetric part of the $N=1$ coefficient combines with the antisymmetric part of the $N=0$ coefficient. Explicitly,
\beq
\begin{aligned}
|\mathbf1]^{\dot\beta}|\mathbf2]^{\dot\gamma}
\left(\bar g^{(0)}_{\dot\beta\dot\gamma}
+p_1^{\dot\alpha\alpha}\bar g^{(1)}_{\dot\beta\dot\gamma;\alpha\dot\alpha}\right)
=&\ |\mathbf1]^{\dot\beta}|\mathbf2]^{\dot\gamma}
\bar g^{(0)}_{(\dot\beta\dot\gamma)}
-[\mathbf{12}]q^{\dot\alpha\alpha}\bar g_{\alpha\dot\alpha}~,\\
\bra{\mathbf1}^{\beta}\bra{\mathbf2}^{\gamma}
\left(g^{(0)}_{\beta\gamma}
+p_1^{\dot\alpha\alpha}g^{(1)}_{\beta\gamma;\alpha\dot\alpha}\right)
=&\ \bra{\mathbf1}^{\beta}\bra{\mathbf2}^{\gamma}
g^{(0)}_{(\beta\gamma)}
+\braket{\mathbf{12}}q^{\dot\alpha\alpha}g_{\alpha\dot\alpha}~.
\end{aligned}
\eeq
For the mixed-chirality bilinears, the massive Dirac equations give the exact transversality identity
\beq
P_{\beta\dot\beta}
\left(
\bra{\mathbf1}^{\beta}|\mathbf2]^{\dot\beta}
+|\mathbf1]^{\dot\beta}\bra{\mathbf2}^{\beta}
\right)=0~.
\eeq
The transverse sum is the spin-triplet covariant $S^\mu$. Decomposing the two coefficient bispinors into transverse and longitudinal parts, their duplicated longitudinal component enters with coefficient $P^{\dot\beta\beta}(h^{(0)}_{\beta\dot\beta}-\bar h^{(0)}_{\dot\beta\beta})$. A nonredundant representative therefore obeys
\beq
P^{\dot\beta\beta}
\left(h^{(0)}_{\beta\dot\beta}-\bar h^{(0)}_{\dot\beta\beta}\right)=0~.
\eeq
Conversely, the retained $N=0$ and $N=1$ structures reconstruct the four covariants in Eq.~(\ref{eq: exact fermion covariants}), so this choice neither omits a $j=1$ sector nor introduces a redundant higher-power coefficient. These are the relations used in the main text. They follow from the exact $SO(3)$ decomposition and algebraic spinor reduction, rather than from the conditional $m\to0$ check in Appendix~\ref{app: fermion Casimir}.

\subsection{Independent multiplicity count and complete all-fermion basis}\label{app: multiplicity basis}

The relative-momentum construction gives four copies of the spin-one representation for each fermion pair. We order them as
\beq
\mathfrak A=(10,01,11,21)~.
\eeq
The complete all-fermion basis in Eq.~(\ref{eq: all fermion covariant basis}) is therefore the following $4\times4$ array:
\beq\label{eq: sixteen basis matrix}
\begin{array}{c|cccc}
 & \widehat T_{10} & \widehat T_{01} & \widehat T_{11} & \widehat T_{21}\\ \hline
T_{10} & T_{10}\!\cdot\!\widehat T_{10} & T_{10}\!\cdot\!\widehat T_{01} & T_{10}\!\cdot\!\widehat T_{11} & T_{10}\!\cdot\!\widehat T_{21}\\
T_{01} & T_{01}\!\cdot\!\widehat T_{10} & T_{01}\!\cdot\!\widehat T_{01} & T_{01}\!\cdot\!\widehat T_{11} & T_{01}\!\cdot\!\widehat T_{21}\\
T_{11} & T_{11}\!\cdot\!\widehat T_{10} & T_{11}\!\cdot\!\widehat T_{01} & T_{11}\!\cdot\!\widehat T_{11} & T_{11}\!\cdot\!\widehat T_{21}\\
T_{21} & T_{21}\!\cdot\!\widehat T_{10} & T_{21}\!\cdot\!\widehat T_{01} & T_{21}\!\cdot\!\widehat T_{11} & T_{21}\!\cdot\!\widehat T_{21}
\end{array}~.
\eeq
Each row and column labels an inequivalent multiplicity sector before the two spin-one indices are contracted. The four sectors are distinguished by the simultaneous $(L^2,S^2)$ eigenvalues $(2,0)$, $(0,2)$, $(2,2)$, and $(6,2)$, so their representatives are linearly independent whenever they are nonzero. Schur's lemma therefore assigns one independent scalar coefficient to every entry. This gives sixteen structures without relying on a choice of spinor coordinates. Reducing the covariants with the massive Dirac equations and Schouten identities reproduces the $N=0,1$ spinor form in Eq.~(\ref{eq: fermion j1 reduced basis}). Conversely, Eq.~(\ref{eq: exact fermion covariants}) reconstructs every row and column from that reduced form, so the two descriptions have the same rank.

\subsection{Optional verification: direct Casimir reduction in spinor variables}\label{app: fermion Casimir}

This subsection is an independent algebraic verification and may be skipped without affecting the proof. The main text uses the exact relative-momentum decomposition of Appendix~\ref{app: fermion detail}. The direct Pauli--Lubanski calculation below reproduces both the four allowed $(L,S)$ sectors and the coefficient restrictions in Eq.~(\ref{eq: many h=g=0}), and hence gives the same reduced basis as Eq.~(\ref{eq: fermion j1 reduced basis}). It is algebraically longer, but it checks the result directly in the spinor form of $W^2$.

Substituting Eq.~(\ref{eq: fermion ansatz}) into Eq.~(\ref{eq: W^2 eigen}) for $j=1$ in the $\mathcal{I}=\{1,2\}$ channel, and using $P_{\alpha\dot{\alpha}}=\varepsilon_{\alpha\beta}\varepsilon_{\dot{\alpha}\dot{\beta}}(P^T)^{\dot{\beta}\beta}$, we obtain
\beq\label{eq: fermion sum=0}
\begin{aligned}
    &\sum_{N\geq 0}\Bigg(\bigg\{Ns\Big(\bra{\mathbf{1}}^{\beta}\bra{\mathbf{2}}^{\gamma}g^N_{\beta\gamma}+|\mathbf{1}]^{\dot{\beta}}|\mathbf{2}]^{\dot{\gamma}}\bar{g}^N_{\dot{\beta}\dot{\gamma}}+\bra{\mathbf{1}}^{\beta}|\mathbf{2}]^{\dot{\beta}}h^N_{\beta\dot{\beta}}+|\mathbf{1}]^{\dot{\beta}}\bra{\mathbf{2}}^{\beta}\bar{h}^N_{\dot{\beta}\beta}\Big)_{;\{\alpha,\dot{\alpha}\}}p_1^{\dot{\alpha}_1\alpha_1}\\
    &+\Big(s\big(-\braket{\mathbf{12}}\varepsilon^{\beta\gamma}g^N_{\beta\gamma}+[\mathbf{12}]\varepsilon^{\dot{\beta}\dot{\gamma}}\bar{g}^N_{\dot{\beta}\dot{\gamma}}\big)+m\big(\braket{\mathbf{12}}-[\mathbf{12}]\big)(P^T)^{\dot{\beta}\beta}\big(h^N_{\beta\dot{\beta}}-\bar{h}^N_{\dot{\beta}\beta}\big)\Big)_{;\{\alpha,\dot{\alpha}\}}p_1^{\dot{\alpha}_1\alpha_1}\\
    &+N\Big[-s\Big(\big(\bra{\mathbf{1}}^\beta\bra{\mathbf{2}}^{\alpha_1}p_1^{\dot{\alpha}_1\gamma}+\bra{\mathbf{1}}^{\alpha_1}\bra{\mathbf{2}}^{\gamma}p_1^{\dot{\alpha}_1\beta}\big)g^N_{\beta\gamma}\\
    &\hspace{2cm}+\big(\bra{\mathbf{1}}^\beta|\mathbf{2}]^{\dot{\alpha}_1}p_1^{\dot{\gamma}\alpha_1}+\bra{\mathbf{1}}^{\alpha_1}|\mathbf{2}]^{\dot{\gamma}}p_1^{\dot{\alpha}_1\beta}\big)h^N_{\beta\dot{\gamma}}\Big)_{;\{\alpha,\dot{\alpha}\}}\\
    &\hspace{1.1cm}+\frac{1}{2}\Big(\big((P^T\ket{\mathbf{2}})^{\dot{\alpha}_1}(Pp_1)^{\alpha_1}_\sigma\varepsilon^{\sigma\gamma}+(\bra{\mathbf{2}}Pp_1)^{\alpha_1}(P^T)^{\dot{
    \alpha}_1\gamma}\big)\big(\bra{\mathbf{1}}^\beta g^N_{\beta\gamma}+|\mathbf{1}]^{\dot{\beta}}\bar{h}^N_{\dot{\beta}\gamma}\big)_{;\{\alpha,\dot{\alpha}\}}\\
    &\hspace{1.7cm}+\big((P^T\ket{\mathbf{1}})^{\dot{\alpha}_1}(Pp_1)^{\alpha_1}_\sigma\varepsilon^{\sigma\gamma}+(\bra{\mathbf{1}}Pp_1)^{\alpha_1}(P^T)^{\dot{
    \alpha}_1\gamma}\big)\big(\bra{\mathbf{2}}^\beta g^N_{\gamma\beta}+|\mathbf{2}]^{\dot{\beta}}h^N_{\gamma\dot{\beta}}\big)_{;\{\alpha,\dot{\alpha}\}}\Big)\\
    &\hspace{1cm}+\Big(\bra{\mathbf{1}}^{\sigma}\leftrightarrow|\mathbf{1}]^{\dot{\sigma}},\bra{\mathbf{2}}^{\sigma}\leftrightarrow|\mathbf{2}]^{\dot{\sigma}},g^N_{\beta\gamma}\leftrightarrow -\bar{g}^N_{\dot{\beta}\dot{\gamma}},h^N_{\beta\dot{\gamma}}\leftrightarrow -\bar{h}^N_{\dot{\beta}\gamma},\varepsilon^{\sigma\rho}\leftrightarrow\varepsilon^{\dot{\sigma}\dot{\rho}}\Big)\Big]\bigg\}\prod_{i=2}^N p_1^{\dot{\alpha}_i\alpha_i}\\
    &\hspace{0.9cm}+\Big(\bra{\mathbf{1}}^{\beta}\bra{\mathbf{2}}^{\gamma}g^N_{\beta\gamma}+|\mathbf{1}]^{\dot{\beta}}|\mathbf{2}]^{\dot{\gamma}}\bar{g}^N_{\dot{\beta}\dot{\gamma}}+\bra{\mathbf{1}}^{\beta}|\mathbf{2}]^{\dot{\beta}}h^N_{\beta\dot{\beta}}+|\mathbf{1}]^{\dot{\beta}}\bra{\mathbf{2}}^{\beta}\bar{h}^N_{\dot{\beta}\beta}\Big)_{;\{\alpha,\dot{\alpha}\}}W^2\prod_{i=1}^N p_1^{\dot{\alpha}_i\alpha_i}\Bigg)\\
    &\hspace{14cm}=0~.
\end{aligned}
\eeq
Note that after swapping $\bra{\mathbf{1}}^{\sigma}\leftrightarrow|\mathbf{1}]^{\dot{\sigma}}$ and $\bra{\mathbf{2}}^{\sigma}\leftrightarrow|\mathbf{2}]^{\dot{\sigma}}$, $p_1^{\dot{\alpha}\beta}\leftrightarrow -p_1^{\dot{\beta}\alpha}$ and $(P^T)^{\dot{\alpha}\beta}\leftrightarrow -(P^T)^{\dot{\beta}\alpha}$.

The exact decomposition restricts the coefficients as in Eq.~(\ref{eq: many h=g=0}). Using these relations and the identities in Appendix~\ref{app: formula}, the Casimir equation reduces as follows.

Using Eq.~(\ref{eq: many h=g=0}) and the identities in Appendix~\ref{app: formula}, we reduce Eq.~(\ref{eq: fermion sum=0}) to
\beq\label{eq: simplified fermion sum}
\begin{aligned}
    &-s\big(\braket{\mathbf{12}}\varepsilon^{\beta\gamma}g^{(N=0)}_{\beta\gamma}-[\mathbf{12}]\varepsilon^{\dot{\beta}\dot{\gamma}}\bar{g}^{(N=0)}_{\dot{\beta}\dot{\gamma}}\big)\\
    &+\bigg[-s\big(2\bra{\mathbf{1}}^\beta\bra{\mathbf{2}}^\gamma+\braket{\mathbf{12}}\varepsilon^{\beta\gamma}\big)p_1^{\dot{\alpha}\alpha}+s\bra{\mathbf{1}}^\beta\bra{\mathbf{2}}^\gamma(P^T)^{\dot{\alpha}\alpha}\\
    &\hspace{0.7cm}+\frac{1}{2}\Big(\bra{\mathbf{1}}^\beta\big((P^T)^{\dot{\rho}\beta}\varepsilon_{\dot{\rho}\dot{\sigma}}p_1^{\dot{\sigma}\alpha}(P^T\ket{\mathbf{2}})^{\dot{\alpha}}+(P^T)^{\dot{\alpha}\gamma}(\bra{\mathbf{2}}Pp_1)^\alpha\big)\\
    &\hspace{1.7cm}+\bra{\mathbf{2}}^\gamma\big((P^T)^{\dot{\rho}\beta}\varepsilon_{\dot{\rho}\dot{\sigma}}p_1^{\dot{\sigma}\alpha}(P^T\ket{\mathbf{1}})^{\dot{\alpha}}+(P^T)^{\dot{\alpha}\beta}(\bra{\mathbf{1}}Pp_1)^\alpha\big)\Big)\bigg]g^{(N=1)}_{\beta\gamma;\alpha\dot{\alpha}}\\
    &+\bigg[-s\big(2|\mathbf{1}]^{\dot{\beta}}|\mathbf{2}]^{\dot{\gamma}}-[\mathbf{12}]\varepsilon^{\dot{\beta}\dot{\gamma}}\big)p_1^{\dot{\alpha}\alpha}+s|\mathbf{1}]^{\dot{\beta}}|\mathbf{2}]^{\dot{\gamma}}(P^T)^{\dot{\alpha}\alpha}\\
    &\hspace{0.7cm}+\frac{1}{2}\Big(|\mathbf{1}]^{\dot{\beta}}\big((P^T)^{\dot{\gamma}\rho}\varepsilon_{\rho\sigma}p_1^{\dot{\alpha}\sigma}([\mathbf{2}|P^T)^\alpha+(P^T)^{\dot{\gamma}\alpha}(p_1P|\mathbf{2}])^{\dot{\alpha}}\big)\\
    &\hspace{1.7cm}+|\mathbf{2}]^{\dot{\gamma}}\big((P^T)^{\dot{\beta}\rho}\varepsilon_{\rho\sigma}p_1^{\dot{\alpha}\sigma}([\mathbf{1}|P^T)^\alpha+(P^T)^{\dot{\beta}\alpha}(p_1P|\mathbf{1}])^{\dot{\alpha}}\big)\Big)\bigg]\bar{g}^{(N=1)}_{\dot{\beta}\dot{\gamma};\alpha\dot{\alpha}}\\
    &+m\big(\braket{\mathbf{12}}-[\mathbf{12}]\big)(P^T)^{\dot{\beta}\beta}\Big(h^{(N=0)}_{\beta\dot{\beta}}-\bar{h}^{(N=0)}_{\dot{\beta}\beta}\Big)=0~.
\end{aligned}
\eeq

Contracting Eq.~(\ref{eq: simplified fermion sum}) with $\braket{1_I2_J}$ and using the identities in Appendix~\ref{app: formula}, we obtain
\beq\label{eq: fermion reduced start}
\begin{aligned}
    &s\bigg\{\varepsilon^{\dot{\beta}\dot{\gamma}}\bar{g}^{(N=0)}_{\dot{\beta}\dot{\gamma}}+\frac{1}{2}\Big((P^T)^{\dot{\alpha}\alpha}\varepsilon^{\dot{\beta}\dot{\gamma}}+(P^T)^{\dot{\gamma}\alpha}\varepsilon^{\dot{\alpha}\dot{\beta}}+(P^T)^{\dot{\beta}\alpha}\varepsilon^{\dot{\alpha}\dot{\gamma}}\Big)\bar{g}^{(N=1)}_{\dot{\beta}\dot{\gamma};\alpha\dot{\alpha}}\bigg\}\\
    -&\bigg\{\frac{1}{2}s\Big(p_1^{\dot{\gamma}\alpha}\varepsilon^{\dot{\alpha}\dot{\beta}}+p_1^{\dot{\beta}\alpha}\varepsilon^{\dot{\alpha}\dot{\gamma}}\Big)\bar{g}^{(N=1)}_{\dot{\beta}\dot{\gamma};\alpha\dot{\alpha}}+m(P^T)^{\dot{\beta}\beta}\Big(h^{(N=0)}_{\beta\dot{\beta}}-\bar{h}^{(N=0)}_{\dot{\beta}\beta}\Big)\bigg\}\\
    +&\bigg\{m^2\Big(2\big(\varepsilon
    ^{\beta\gamma}g^{(N=0)}_{\beta\gamma}-\varepsilon^{\dot{\beta}\dot{\gamma}}\bar{g}^{(N=0)}_{\dot{\beta}\dot{\gamma}}\big)+(P^T)^{\dot{\alpha}\alpha}\varepsilon^{\beta\gamma}g^{(N=1)}_{\beta\gamma;\dot{\alpha}\alpha}\\
    &\hspace{1cm}-\big((P^T)^{\dot{\alpha}\alpha}\varepsilon^{\dot{\beta}\dot{\gamma}}+(P^T)^{\dot{\gamma}\alpha}\varepsilon^{\dot{\alpha}\dot{\beta}}+(P^T)^{\dot{\beta}\alpha}\varepsilon^{\dot{\alpha}\dot{\gamma}}\big)\bar{g}^{(N=1)}_{\dot{\beta}\dot{\gamma};\alpha\dot{\alpha}}\Big)\\
    &\hspace{0.2cm}+2\Big((P^Tp_1)^{\dot{\gamma}}_{\dot{\sigma}}\varepsilon^{\dot{\sigma}\dot{\beta}}p_1^{\dot{\alpha}\alpha}+(\dot{\beta}\leftrightarrow\dot{\gamma})\Big)\bar{g}^{(N=1)}_{\dot{\beta}\dot{\gamma};\alpha\dot{\alpha}}\bigg\}=0~,
\end{aligned}
\eeq
The three curly brackets are the independent transverse-harmonic projections obtained from the exact decomposition in Appendix~\ref{app: fermion detail}. Since harmonics of different rank belong to inequivalent $SO(3)$ representations, each bracket must vanish separately.
\beq\label{eq: fermion reduced step 0}
\varepsilon^{\dot{\beta}\dot{\gamma}}\bar{g}^{(N=0)}_{\dot{\beta}\dot{\gamma}}+\frac{1}{2}\Big((P^T)^{\dot{\alpha}\alpha}\varepsilon^{\dot{\beta}\dot{\gamma}}+(P^T)^{\dot{\gamma}\alpha}\varepsilon^{\dot{\alpha}\dot{\beta}}+(P^T)^{\dot{\beta}\alpha}\varepsilon^{\dot{\alpha}\dot{\gamma}}\Big)\bar{g}^{(N=1)}_{\dot{\beta}\dot{\gamma};\alpha\dot{\alpha}}=0~.
\eeq
\beq\label{eq: fermion reduced step 0.5}
\frac{1}{2}s\Big(p_1^{\dot{\gamma}\alpha}\varepsilon^{\dot{\alpha}\dot{\beta}}+p_1^{\dot{\beta}\alpha}\varepsilon^{\dot{\alpha}\dot{\gamma}}\Big)\bar{g}^{(N=1)}_{\dot{\beta}\dot{\gamma};\alpha\dot{\alpha}}+m(P^T)^{\dot{\beta}\beta}\Big(h^{(N=0)}_{\beta\dot{\beta}}-\bar{h}^{(N=0)}_{\dot{\beta}\beta}\Big)=0
\eeq
\beq\label{eq: fermion reduced step 1}
\begin{aligned}
    &m^2\Big(2\big(\varepsilon
    ^{\beta\gamma}g^{(N=0)}_{\beta\gamma}-\varepsilon^{\dot{\beta}\dot{\gamma}}\bar{g}^{(N=0)}_{\dot{\beta}\dot{\gamma}}\big)+(P^T)^{\dot{\alpha}\alpha}\varepsilon^{\beta\gamma}g^{(N=1)}_{\beta\gamma;\dot{\alpha}\alpha}\\
    &\hspace{2cm}-\big((P^T)^{\dot{\alpha}\alpha}\varepsilon^{\dot{\beta}\dot{\gamma}}+(P^T)^{\dot{\gamma}\alpha}\varepsilon^{\dot{\alpha}\dot{\beta}}+(P^T)^{\dot{\beta}\alpha}\varepsilon^{\dot{\alpha}\dot{\gamma}}\big)\bar{g}^{(N=1)}_{\dot{\beta}\dot{\gamma};\alpha\dot{\alpha}}\Big)\\
    =&-2\Big((P^Tp_1)^{\dot{\gamma}}_{\dot{\sigma}}\varepsilon^{\dot{\sigma}\dot{\beta}}p_1^{\dot{\alpha}\alpha}+(\dot{\beta}\leftrightarrow\dot{\gamma})\Big)\bar{g}^{(N=1)}_{\dot{\beta}\dot{\gamma};\alpha\dot{\alpha}}~.
\end{aligned}
\eeq

For a further conditional check, assume that the coefficient functions are regular at $m=0$. This assumption is not used in the main basis derivation. Taking $m\rightarrow0$ in Eq.~(\ref{eq: fermion reduced step 1}) gives
\beq
\Big[(P^Tp_1)^{\dot{\gamma}}_{\dot{\sigma}}\varepsilon^{\dot{\sigma}\dot{\beta}}p_1^{\dot{\alpha}\alpha}+(\dot{\beta}\leftrightarrow\dot{\gamma})\Big]\bar{g}^{(N=1)}_{\dot{\beta}\dot{\gamma};\alpha\dot{\alpha}}=0~.
\eeq
The terms in $[\cdots]$ do not vanish in general and are fully symmetric in $\dot{\beta}\leftrightarrow\dot{\gamma}$. Thus
\beq\label{eq: g1 antisymmetric}
\bar{g}^{(N=1)}_{\dot{\beta}\dot{\gamma};\alpha\dot{\alpha}}=-\bar{g}^{(N=1)}_{\dot{\gamma}\dot{\beta};\alpha\dot{\alpha}}~.
\eeq
Applying Eq.~(\ref{eq: g1 antisymmetric}) to Eq.~(\ref{eq: fermion reduced step 0}) gives
\beq\label{eq: fermion reduced step 2}
\varepsilon^{\dot{\beta}\dot{\gamma}}\bar{g}^{(N=0)}_{\dot{\beta}\dot{\gamma}}+\frac{1}{2}(P^T)^{\dot{\alpha}\alpha}\varepsilon^{\dot{\beta}\dot{\gamma}}\bar{g}^{(N=1)}_{\dot{\beta}\dot{\gamma};\alpha\dot{\alpha}}=0~.
\eeq
Since $\bar{g}^{(N=1)}_{\dot{\beta}\dot{\gamma};\alpha\dot{\alpha}}$ is antisymmetric under $\dot{\beta}\leftrightarrow\dot{\gamma}$, we can write it as
\beq
\bar{g}^{(N=1)}_{\dot{\beta}\dot{\gamma};\alpha\dot{\alpha}}\equiv\bar{g}_{\alpha\dot{\alpha}}\varepsilon_{\dot{\beta}\dot{\gamma}}~,
\eeq
where $\bar{g}_{\alpha\dot{\alpha}}=\bar{g}(p_3,p_4)_{\alpha\dot{\alpha}}$ is a function of $p_3^\mu$ and $p_4^\mu$. We parameterize the antisymmetric part of $\bar{g}^{(N=0)}$ as
\beq
\bar{g}_{\text{asym}}\varepsilon_{\dot{\beta}\dot{\gamma}}\equiv\bar{g}^{(N=0)}_{\dot{\beta}\dot{\gamma}}-\bar{g}^{(N=0)}_{\dot{\gamma}\dot{\beta}}~,
\eeq
Equation~(\ref{eq: fermion reduced step 2}) becomes
\beq
\varepsilon^{\dot{\beta}\dot{\gamma}}\frac{1}{2}\Big(\bar{g}^{(N=0)}_{\dot{\beta}\dot{\gamma}}-\bar{g}^{(N=0)}_{\dot{\gamma}\dot{\beta}}\Big)+\frac{1}{2}(P^T)^{\dot{\alpha}\alpha}\varepsilon^{\dot{\beta}\dot{\gamma}}\bar{g}^{(N=1)}_{\dot{\beta}\dot{\gamma};\alpha\dot{\alpha}}=-\bar{g}_{\text{asym}}-(P^T)^{\dot{\alpha}\alpha}\bar{g}_{\alpha\dot{\alpha}}=0~.
\eeq
Therefore, $\bar{g}_{\text{asym}}=-(P^T)^{\dot{\alpha}\alpha}\bar{g}_{\alpha\dot{\alpha}}$, and we obtain
\beq\label{eq: fermion reduced end}
\begin{aligned}
    |\mathbf{1}]^{\dot{\beta}}|\mathbf{2}]^{\dot{\gamma}}\Big(\bar{g}^{(N=0)}_{\dot{\beta}\dot{\gamma}}+p_1^{\dot{\alpha}\alpha}\bar{g}^{(N=1)}_{\dot{\beta}\dot{\gamma};\alpha\dot{\alpha}}\Big)=&|\mathbf{1}]^{\dot{\beta}}|\mathbf{2}]^{\dot{\gamma}}\frac{1}{2}\Bigl(\bar{g}^{(N=0)}_{\dot{\beta}\dot{\gamma}}+\bar{g}^{(N=0)}_{\dot{\gamma}\dot{\beta}}\Bigr)\\
    &-[\mathbf{12}]\frac{1}{2}\bigl(p_1-p_2\bigr)^{\dot{\alpha}\alpha}\bar{g}_{\alpha\dot{\alpha}}~.
\end{aligned}
\eeq

If instead we contract Eq.~(\ref{eq: simplified fermion sum}) with $[1_I2_J]$ and follow the steps from Eq.~(\ref{eq: fermion reduced start}) through Eq.~(\ref{eq: fermion reduced end}), we obtain
\beq\label{eq: fermion reduced end 2}
\begin{aligned}
    \bra{\mathbf{1}}^{\beta}\bra{\mathbf{2}}^{\gamma}\Big(g^{(N=0)}_{\beta\gamma}+p_1^{\dot{\alpha}\alpha}g^{(N=1)}_{\beta\gamma;\alpha\dot{\alpha}}\Big)=&\bra{\mathbf{1}}^{\beta}\bra{\mathbf{2}}^{\gamma}\frac{1}{2}\Bigl(g^{(N=0)}_{\beta\gamma}+g^{(N=0)}_{\gamma\beta}\Bigr)\\
    &+\braket{\mathbf{12}}\frac{1}{2}\bigl(p_1-p_2\bigr)^{\dot{\alpha}\alpha}g_{\alpha\dot{\alpha}}~,
\end{aligned}
\eeq
where $g_{\alpha\dot{\alpha}}\varepsilon_{\beta\gamma}\equiv g^{(N=1)}_{\beta\gamma;\alpha\dot{\alpha}}$. On the other hand, substituting Eq.~(\ref{eq: g1 antisymmetric}) into Eq.~(\ref{eq: fermion reduced step 0.5}) gives
\beq\label{eq: fermion h constraint}
(P^T)^{\dot{\beta}\beta}\big(h^{(N=0)}_{\beta\dot{\beta}}-\bar{h}^{(N=0)}_{\dot{\beta}\beta}\big)=0~.
\eeq
Under the stated regularity assumption, these relations reproduce the reduced basis in Eq.~(\ref{eq: fermion j1 reduced basis}); the basis itself was obtained independently from the exact relative-momentum decomposition.

\section{Finite-mass fermion currents and the transverse helicity representative}\label{app: exact minimal pole}

This appendix derives Eq.~(\ref{eq: exact vector current LS}) and the transverse representative in Eqs.~(\ref{eq: exact axial current LS})--(\ref{eq: axial vector helicity relation}) without taking a high-energy limit. Consider a fermion--antifermion pair of mass $m$ in its center-of-momentum frame,
\beq
P^\mu=(2E,\mathbf0)~,
\qquad
p_1^\mu=(E,\mathbf q)~,
\qquad
p_2^\mu=(E,-\mathbf q)~,
\qquad
E=\frac{\sqrt{s}}{2}~,
\eeq
with $\mathbf q^2=E^2-m^2$. Choose Dirac spinors
\beq
u(p_1)=\sqrt{E+m}
\begin{pmatrix}
\xi\\[1mm]
\dfrac{\boldsymbol\sigma\cdot\mathbf q}{E+m}\xi
\end{pmatrix}~,
\qquad
v(p_2)=\sqrt{E+m}
\begin{pmatrix}
-\dfrac{\boldsymbol\sigma\cdot\mathbf q}{E+m}\eta\\[1mm]
\eta
\end{pmatrix}~.
\eeq
The two-particle spin singlet and triplet are represented by
\beq
\Sigma=\xi^\dagger\eta~,
\qquad
\mathbf S=\xi^\dagger\boldsymbol\sigma\eta~.
\eeq

For the vector current $V_m^\mu=\overline u\gamma^\mu v$, direct multiplication gives
\beq
V_m^0=0~,
\eeq
and
\beq
(V_m)^i=(E+m)\,\xi^\dagger
\left[\sigma^i-
\frac{(\boldsymbol\sigma\cdot\mathbf q)\sigma^i
(\boldsymbol\sigma\cdot\mathbf q)}{(E+m)^2}
\right]\eta~.
\eeq
Using
\beq
(\boldsymbol\sigma\cdot\mathbf q)\sigma^i
(\boldsymbol\sigma\cdot\mathbf q)
=2q^i(\boldsymbol\sigma\cdot\mathbf q)-\mathbf q^2\sigma^i~,
\eeq
we find
\beq\label{eq: appendix exact vector current}
\mathbf V_m
=2E\,\mathbf S-
\frac{2}{E+m}\mathbf q(\mathbf q\cdot\mathbf S)~.
\eeq
In the same frame, the spatial parts of the raw covariants in Eq.~(\ref{eq: exact fermion covariants}) are
\beq
\mathbf T_{01}=\mathbf S~,
\qquad
\mathbf T_{21}=-\mathbf q(\mathbf q\cdot\mathbf S)
+\frac{\mathbf q^2}{3}\mathbf S~.
\eeq
Equation~(\ref{eq: appendix exact vector current}) therefore becomes
\beq
V_m^\mu
=\frac{2}{3}(2E+m)T_{01}^\mu
+\frac{2}{E+m}T_{21}^\mu~,
\eeq
which is Eq.~(\ref{eq: exact vector current LS}) after using $2E=\sqrt{s}$. The vector current contains only the ${}^3S_1$ and ${}^3D_1$ sectors; their relative coefficient is fixed by the dimension-four vertex.

The axial current $A_m^\mu=\overline u\gamma^\mu\gamma^5v$ gives
\beq
A_m^0=2m\Sigma~,
\qquad
\mathbf A_m=-2i\,\mathbf q\times\mathbf S~.
\eeq
The time component is longitudinal to $P^\mu$ and is removed by the transverse projection defined in Eq.~(\ref{eq: exact axial current LS}). Since in the pair rest frame
\beq
\mathbf T_{11}=\mathbf q\times\mathbf S
\eeq
up to the sign fixed by the Levi-Civita convention, the transverse axial current belongs entirely to the ${}^3P_1$ sector,
\beq
A_{m\perp}^\mu=-2iT_{11}^\mu~,
\eeq
with the same convention.

To relate this result to the magnetic photon-helicity phase, take the circular polarization vectors about the $\mathbf q$ axis to obey
\beq
\mathbf q\times\boldsymbol\epsilon_h=-ih|\mathbf q|\boldsymbol\epsilon_h~.
\eeq
The transverse projections of the two currents then satisfy
\beq
\epsilon_h\cdot A_{m\perp}
=h\frac{|\mathbf q|}{E}\epsilon_h\cdot V_m
=h\,\beta_m\,\epsilon_h\cdot V_m~,
\eeq
up to a common phase and the choice of whether the polarization is treated as incoming or outgoing. The relative factor $h$ is invariant. Hence the helicity-dependent sign distinguishing the minimal magnetic and electric vertices can be represented on the two physical transverse photon polarizations by $T_{11}$.

This relation does not uniquely determine a vector in the full three-dimensional massive $j=1$ space: a component along the $m_j=0$ direction is invisible to the two circular-polarization contractions. Moreover, the pair rest frame and $\beta_m$ parametrization are singular at the photon factorization locus $s=0$. The $T_{11}={}^3P_1$ current is therefore a transverse representative of the helicity phase, not a unique finite-mass continuation of the pole data. The exact information used in the main proof is instead that the minimal magnetic vertex has the same massive spinor polynomial as the vector vertex and is consequently spin triplet, as stated in Eq.~(\ref{eq: exact minimal magnetic spin sector}).

\section{Review of all electric on-shell gluing}\label{app: all e construct}
To show that the $x$-factor
\beq
x_{12}=\frac{\bra{\zeta}(p_1-p_2)|k]}{2m\braket{\zeta k}}~
\eeq
is independent of $\ket{\zeta}$, we differentiate it with respect to $\bra{\zeta}^\alpha$.
\beq
\begin{aligned}
    &\frac{\partial}{\partial\bra{\zeta}^\alpha}\frac{\bra{\zeta}(p_1-p_2)|k]}{2m\braket{\zeta k}}\\
    =&\frac{\prescript{}{\alpha}{(p_1-p_2)}|k]}{2m\braket{\zeta k}}-\frac{\prescript{}{\alpha}{\ket{k}}\bra{\zeta}(p_1-p_2)|k]}{2m\braket{\zeta k}^2}\\
    =&-\frac{\prescript{}{\alpha}{(p_1-p_2)}|k]\braket{k\zeta}}{2m\braket{\zeta k}^2}-\frac{\prescript{}{\alpha}{\ket{k}}[k|(p_1-p_2)\ket{\zeta}}{2m\braket{\zeta k}^2}\\
    =&-\frac{\prescript{}{\alpha}{\left[(p_1-p_2)k+k(p_1-p_2)\right]}\ket{\zeta}}{2m\braket{\zeta k}^2}\\
    =&-2\left[(p_1-p_2)\cdot k\right]\frac{\prescript{}{\alpha}{\ket{\zeta}}}{2m\braket{\zeta k}^2}\\
    =&0~,
\end{aligned}
\eeq
where we used $(p_1-p_2)\cdot k=0$ at the end.

To calculate the all-electric residue by on-shell gluing, note that
\beq
\frac{1}{x_{12}}=\frac{\bra{k}(p_1-p_2)|\zeta]}{2m[\zeta k]}~.
\eeq
For scalar-to-scalar production,
\beq
\begin{aligned}
    &m_\phi m_{\phi'}\frac{x_{34}}{x_{12}}+m_\phi m_{\phi'} \frac{x_{12}}{x_{34}}\\
    =&-\left(\frac{\bra{k}(p_1-p_2)|\zeta]}{2[\zeta k]}\frac{\bra{\zeta}(p_3-p_4)|k]}{2\braket{\zeta k}}+\frac{\bra{\zeta}(p_1-p_2)|k]}{2\braket{\zeta k}}\frac{\bra{k}(p_3-p_4)|\zeta]}{2[\zeta k]}\right)~,
\end{aligned}
\eeq
where the minus sign follows from the analytic continuation convention for $x_{34}$~\cite{Elvang:2015rqa}, with $\ket{-k}=-\ket{k}$ and $|-k]=+|k]$. Using Eq.~(\ref{eq: J K definition}) and defining $J_s'\equiv(p_3-p_4)/\sqrt{2}$, the first term in parentheses becomes
\beq\label{eq: electric gluing}
\frac{\bra{k}(p_1-p_2)|\zeta]}{2[\zeta k]}\frac{\bra{\zeta}(p_3-p_4)|k]}{2\braket{\zeta k}}= \frac{\bra{k}J_s|\zeta]}{\sqrt{2}[\zeta k]}\frac{\bra{\zeta}J_s'|k]}{\sqrt{2}\braket{\zeta k}}=\frac{\Tr{\left(|k]\bra{k}J_s\zeta J_s'\right)}}{-4(k \cdot\zeta)}~,
\eeq
while the second term becomes
\beq
\frac{\bra{\zeta}(p_1-p_2)|k]}{2\braket{\zeta k}}\frac{\bra{k}(p_3-p_4)|\zeta]}{2[\zeta k]}=\frac{\Tr{\left(\ket{k}[k|J_s\zeta J_s'\right)}}{-4(k \cdot\zeta)}~.
\eeq
Note the distinction between $|k]\bra{k}=k^\mu \bar{\sigma}_\mu$ and $\ket{k}[k|=k^\mu\sigma_\mu$. Using Eq.~(\ref{eq: trace}) and $k\cdot J_s=k\cdot J_s'=0$,
\beq
\begin{aligned}
    &m_\phi m_{\phi '}\frac{x_{34}}{x_{12}}+m_\phi m_{\phi '} \frac{x_{12}}{x_{34}}\\
    =&\frac{\Tr{\left(|k]\bra{k}J_s\zeta J_s'\right)}}{-4(k \cdot\zeta)}+\frac{\Tr{\left(\ket{k}[k|J_s\zeta J_s'\right)}}{-4(k \cdot\zeta)}\\
    =&J_s\cdot J_s'~.
\end{aligned}
\eeq

When fermions are involved, we use the Schouten identity in Eq.~(\ref{eq: schouten}),
\beq
\begin{aligned}
    x_{12}\braket{\mathbf{12}}=&\frac{\bra{\zeta}(p_1-p_2)|k]}{2m\braket{\zeta k}}\braket{\mathbf{12}}\\
    =&-\left(\braket{\mathbf{2}\zeta}\bra{\mathbf{1}}+\braket{\zeta \mathbf{1}}\bra{\mathbf{2}}\right)\frac{(p_1-p_2)|k]}{2m\braket{\zeta k}}\\
=&\frac{\bra{\zeta}\left(\ket{\mathbf{1}}[\mathbf{2}|+\ket{\mathbf{2}}[\mathbf{1}|\right)|k]}{\braket{k\zeta}}\\
=&\frac{\bra{\zeta}J_f|k]}{\sqrt{2}\braket{k\zeta}}~,
\end{aligned}
\eeq
where we used Eq.~(\ref{eq: J K definition}) in the last step. Since $k\cdot J_f=0$ just like $J_s$, we can follow the procedure after Eq.~(\ref{eq: electric gluing}) and find that all-electric residues have the form $J\cdot J'$. All-magnetic residues can be calculated similarly by replacing $J$ with $K$ and carefully canceling the minus sign in Eq.~(\ref{eq: magnetic 3-point}).

\section{Discrete symmetry of electric-magnetic scattering}\label{sec: CP for scattering}

In this section, we repeat the discrete-symmetry analysis for an electric charge scattering from a magnetic charge and explain why the production selection rule does not carry over. We classify symmetry-permitted lowest partial waves only. The known scattering constructions in Refs.~\cite{Csaki:2020inw,Csaki:2020yei} establish the nonzero amplitudes and their dynamical normalization; the tensors below are not by themselves a derivation that a particular three-point normalization factorizes through an exchanged particle. For simplicity, we present only the lowest partial wave.

A scattering amplitude must carry the pairwise little-group weight in Eq.~(\ref{eq: qij}). Opposite pairwise helicities therefore require different spinor expressions. Let the electric charge carry momentum $p_1^\mu$ in the incoming state and $p_3^\mu$ in the outgoing state, and let the monopole carry $p_2^\mu$ and $p_4^\mu$, respectively. The pairwise helicities in the two states are the same, $h_{12}=h_{34}$. Following Refs.~\cite{Csaki:2020inw,Csaki:2020yei}, we write the $j$ partial-wave amplitude as
\beq
\mathcal{B}^j_{h_{12}}(p_1,p_2;p_3,p_4)_{\{h\}}\sim\bra{0}\{a_3^{h_3}\alpha_4^{h_4}\}\hat{S}^j\{a_1^{h_1}\alpha_2^{h_2}\}^{\dagger}\ket{0}~.
\eeq
The pairwise helicity $h_{12}$ is distinct from the ordinary helicities of the individual particles, $h_i$, denoted collectively by $\{h\}$. We grouped $\{a_1^{h_1}\alpha_2^{h_2}\}^{\dagger}$ and $\{a_3^{h_3}\alpha_4^{h_4}\}$ together to emphasize that these multiparticle states are not direct products of single-particle states because they carry pairwise helicity. Under $\mathcal{C}_M\mathcal{P}$, they transform as\footnote{Since the initial and final states contain the same particle types, the intrinsic-parity factors square to 1 at the $S$-matrix level, so we omit them here.}:
\beq
(\mathcal{C}_M\mathcal{P})^{-1}\{a_i^{h_i}\alpha_j^{h_j}\}^{\dagger}\mathcal{C}_M\mathcal{P}=\{a_{\bar{i}}^{(-h_i)}\beta_{\bar{j}}^{(-h_j)}\}^{\dagger}~.
\eeq
Under $\mathcal{C}_E\mathcal{C}_M$, they transform as
\beq
(\mathcal{C}_E\mathcal{C}_M)^{-1}\{a_i^{h_i}\alpha_j^{h_j}\}^{\dagger}\mathcal{C}_E\mathcal{C}_M=\{b_{i}^{(h_i)}\beta_{j}^{(h_j)}\}^{\dagger}~.
\eeq
For simplicity, we suppress the particle-type label on $\mathcal{B}$; the particle content is clear from each subsection.

Applying $\mathcal{C}_E\mathcal{C}_M$ to the $S$ matrix gives
\beq
\bra{0}\{a_3^{h_3}\alpha_4^{h_4}\}\hat{S}^j\{a_1^{h_1}\alpha_2^{h_2}\}^{\dagger}\ket{0}=\bra{0}\{b_3^{h_3}\beta_4^{h_4}\}\hat{S}^j\{b_1^{h_1}\beta_2^{h_2}\}^{\dagger}\ket{0}~.
\eeq
Since both electric and magnetic charges flip sign, the pairwise helicity is unchanged. Therefore, $\mathcal{C}_E\mathcal{C}_M$ only relates $\mathcal B_{h_{12}}^j$ to itself and does not constrain the amplitude. Thus, we discuss only $\mathcal{C}_M\mathcal{P}$ in the rest of this appendix. To better compare with existing literature, we adopt the all-outgoing convention for the spinor-helicity variables, but distinguish the incoming and outgoing states in the $S$ matrix $\bra{0}\cdots\hat{S}\cdots\ket{0}$ to avoid cluttering.

\subsection{Electric scalar and scalar monopole}
For an electric scalar scattering off a scalar monopole, the lowest partial wave is $j=1/2$. For $h_{12}=\pm1/2$, there are two possible spinor contractions with the required little-group covariance.
\beq
\braket{p_{12}^{\flat\pm}p_{34}^{\flat\pm}}\quad\text{or}\quad[p_{12}^{\flat\mp}p_{34}^{\flat\mp}]~.
\eeq
Applying $\mathcal{C}_M\mathcal{P}$ to the $S$ matrix gives
\beq
\bra{0}\{a_3\alpha_4\}\hat{S}\,\{a_1\alpha_2\}^{\dagger}\ket{0}=\bra{0}\{a_{\bar{3}}\beta_{\bar{4}}\}\hat{S}\,\{a_{\bar{1}}\beta_{\bar{2}}\}^{\dagger}\ket{0}~.
\eeq
The right-hand side now corresponds to the opposite pairwise helicity since the magnetic charge flips sign. Symmetry under $\mathcal{C}_M\mathcal{P}$ thus relates $\mathcal{B}_{h_{12}=+1/2}$ to $\mathcal{B}_{h_{12}=-1/2}$.
\beq
\mathcal{B}^{j=\frac{1}{2}}_{h_{12}=\pm\frac{1}{2}}(p_1,p_2;p_3,p_4)=\mathcal{B}^{j=\frac{1}{2}}_{h_{12}=\mp\frac{1}{2}}(\bar{p}_1,\bar{p}_2;\bar{p}_3,\bar{p}_4)~.
\eeq
Therefore, the amplitude with the correct transformation properties is
\beq
\mathcal{B}^{j=\frac{1}{2}}_{h_{12}=\pm\frac{1}{2}}\sim\braket{p_{12}^{\flat\pm}p_{34}^{\flat\pm}}+[p_{12}^{\flat\mp}p_{34}^{\flat\mp}]~,
\eeq
up to a little-group invariant. The scalar and fermionic cases are summarized in Table~\ref{tb: scattering}.

\subsection{Electric scalar and fermionic monopole}
When one charge is scalar and the other is fermionic, the lowest partial wave is $j=0$. For $h_{12}=\pm1/2$ with the monopole taken to be the fermion, the $j=0$ terms with the correct little-group structure are
\beq
\braket{\mathbf{2}p_{12}^{\flat \pm}}\braket{\mathbf{4}p_{34}^{\flat \pm}}~,\quad\braket{\mathbf{2}p_{12}^{\flat \pm}}[\mathbf{4}p_{34}^{\flat \mp}]~,\quad[\mathbf{2}p_{12}^{\flat \mp}]\braket{\mathbf{4}p_{34}^{\flat \pm}}\text{ or}\quad[\mathbf{2}p_{12}^{\flat \mp}][\mathbf{4}p_{34}^{\flat \mp}]~.
\eeq
As in the all-scalar case, we can analyze the $\mathcal{C}_M\mathcal{P}$ constraint on this amplitude. Under $\mathcal{C}_M\mathcal{P}$, the $S$-matrix satisfies
\beq
\bra{0}\{a_3\alpha_4^{h_4}\}\hat{S}\,\{a_1\alpha_2^{h_2}\}^{\dagger}\ket{0}=\bra{0}\{a_{\bar{3}}\beta_{\bar{4}}^{(-h_4)}\}\hat{S}\,\{a_{\bar{1}}\beta_{\bar{2}}^{(-h_2)}\}^{\dagger}\ket{0}~.
\eeq
The right-hand side again corresponds to the opposite pairwise helicity. Therefore, the amplitudes obey
\beq
\mathcal{B}_{h_{12}=\pm\frac{1}{2}}(p_1,p_2;p_3,p_4)_{\{h\}}=\mathcal{B}_{h_{12}=\mp\frac{1}{2}}(\bar{p}_1,\bar{p}_2;\bar{p}_3,\bar{p}_4)_{\{-h\}}~.
\eeq
Minimal coupling makes the opposite-helicity fermion configurations dominant in the high-energy limit. Thus, retaining only the terms that satisfy minimal coupling, the amplitude is
\beq
\mathcal{B}_{h_{12}=\pm\frac{1}{2}}(p_1,p_2;p_3,p_4)_{\{h\}}\sim \left(\braket{\mathbf{2}p_{12}^{\flat \pm}}[\mathbf{4}p_{34}^{\flat \mp}]+[\mathbf{2}p_{12}^{\flat \mp}]\braket{\mathbf{4}p_{34}^{\flat \pm}}\right)\Big|_{\{h\}}~,
\eeq
up to a little-group invariant.

\begin{table}[ht]
\centering
\begin{tabular}[t]{c|c|c}
(electric, magnetic) & Lowest $j$ and $h_{12}$ & $\mathcal{B}^j_{h_{12}}$ $\vphantom{\frac{1}{2_2}}$ \\
\hline
(scalar, scalar) & $ j=1/2$, $h_{12}=\pm1/2$ & $\braket{p_{12}^{\flat\pm}p_{34}^{\flat\pm}}+[p_{12}^{\flat\mp}p_{34}^{\flat\mp}]$ $\vphantom{\frac{{1^1}^1}{2}}$\\
\hline
(scalar, fermion) & $j=0$, $h_{12}=\pm1/2$ & $\braket{\mathbf{2}p_{12}^{\flat \pm}}[\mathbf{4}p_{34}^{\flat \mp}]+[\mathbf{2}p_{12}^{\flat \mp}]\braket{\mathbf{4}p_{34}^{\flat \pm}}$ $\vphantom{\frac{{1^1}^1}{2}}$ \\
\hline
(fermion, fermion) & $j=0$, $h_{12}=\pm1$ & $\big(\braket{\mathbf{1}p_{12}^{\flat \pm}}\braket{\mathbf{2}p_{12}^{\flat \pm}}[\mathbf{3}p_{34}^{\flat \mp}][\mathbf{4}p_{34}^{\flat \mp}]$ $\vphantom{\frac{{1^1}^1}{2}}$\\
& & $+[\mathbf{1}p_{12}^{\flat \mp}][\mathbf{2}p_{12}^{\flat \mp}]\braket{\mathbf{3}p_{34}^{\flat \pm}}\braket{\mathbf{4}p_{34}^{\flat \pm}}\big)$\\
& & or \\
& & $\big(\braket{\mathbf{1}p_{12}^{\flat \pm}}[\mathbf{2}p_{12}^{\flat \mp}][\mathbf{3}p_{34}^{\flat \mp}]\braket{\mathbf{4}p_{34}^{\flat \pm}}$\\
& & $+[\mathbf{1}p_{12}^{\flat \mp}]\braket{\mathbf{2}p_{12}^{\flat \pm}}\braket{\mathbf{3}p_{34}^{\flat \pm}}[\mathbf{4}p_{34}^{\flat \mp}]\big)$
\end{tabular}
\caption{Particle types, lowest $j$ partial wave, and scattering amplitude $\mathcal{B}^j_{h_{12}}$ up to little-group invariants. Pairwise helicity is denoted by $h_{12}$, while the ordinary particle helicities $\{h\}$ are omitted.}
\label{tb: scattering}
\end{table}

\subsection{Electric fermion and fermionic monopole}
When all particles are fermions, the lowest partial wave is $j=0$. For $h_{12}=\pm1$, the terms satisfying little-group covariance and minimal coupling are
\beq
\begin{aligned}
    &\braket{\mathbf{1}p_{12}^{\flat \pm}}\braket{\mathbf{2}p_{12}^{\flat \pm}}[\mathbf{3}p_{34}^{\flat \mp}][\mathbf{4}p_{34}^{\flat \mp}]~,\quad\braket{\mathbf{1}p_{12}^{\flat \pm}}[\mathbf{2}p_{12}^{\flat \mp}][\mathbf{3}p_{34}^{\flat \mp}]\braket{\mathbf{4}p_{34}^{\flat \pm}}\\
    &[\mathbf{1}p_{12}^{\flat \mp}]\braket{\mathbf{2}p_{12}^{\flat \pm}}\braket{\mathbf{3}p_{34}^{\flat \pm}}[\mathbf{4}p_{34}^{\flat \mp}]~\text{ or}\quad[\mathbf{1}p_{12}^{\flat \mp}][\mathbf{2}p_{12}^{\flat \mp}]\braket{\mathbf{3}p_{34}^{\flat \pm}}\braket{\mathbf{4}p_{34}^{\flat \pm}}~.
\end{aligned}
\eeq
Applying $\mathcal{C}_M\mathcal{P}$ to the $S$ matrix gives
\beq
\bra{0}\{a_3^{h_3}\alpha_4^{h_4}\}\hat{S}\,\{a_1^{h_1}\alpha_2^{h_2}\}^{\dagger}\ket{0}=\bra{0}\{a_{\bar{3}}^{(-h_3)}\beta_{\bar{4}}^{(-h_4)}\}\hat{S}\,\{a_{\bar{1}}^{(-h_1)}\beta_{\bar{2}}^{(-h_2)}\}^{\dagger}\ket{0}~.
\eeq
Therefore, the amplitudes obey
\beq
\mathcal{B}_{h_{12}=\pm1}(p_1,p_2;p_3,p_4)_{\{h\}}=\mathcal{B}_{h_{12}=\mp1}(\bar{p}_1,\bar{p}_2;\bar{p}_3,\bar{p}_4)_{\{-h\}}~.
\eeq
The two combinations satisfying $\mathcal{C}_M\mathcal{P}$ are
\beq
\mathcal{B}_{h_{12}=\pm1}(p_1,p_2;p_3,p_4)_{\{h\}}\sim\begin{cases}
    \left(\braket{\mathbf{1}p_{12}^{\flat \pm}}\braket{\mathbf{2}p_{12}^{\flat \pm}}[\mathbf{3}p_{34}^{\flat \mp}][\mathbf{4}p_{34}^{\flat \mp}] + [\mathbf{1}p_{12}^{\flat \mp}][\mathbf{2}p_{12}^{\flat \mp}]\braket{\mathbf{3}p_{34}^{\flat \pm}}\braket{\mathbf{4}p_{34}^{\flat \pm}}\right)\Big|_{\{h\}}\\
    \hspace{4cm}\text{or}\\
    \left(\braket{\mathbf{1}p_{12}^{\flat \pm}}[\mathbf{2}p_{12}^{\flat \mp}][\mathbf{3}p_{34}^{\flat \mp}]\braket{\mathbf{4}p_{34}^{\flat \pm}}+[\mathbf{1}p_{12}^{\flat \mp}]\braket{\mathbf{2}p_{12}^{\flat \pm}}\braket{\mathbf{3}p_{34}^{\flat \pm}}[\mathbf{4}p_{34}^{\flat \mp}]\right)\Big|_{\{h\}}
\end{cases}
\eeq
up to little-group invariants.

Thus $\mathcal{C}_M\mathcal{P}$, $\mathcal{C}_E\mathcal{C}_M$, angular momentum conservation, pairwise little-group covariance, and minimal coupling do not impose the production-channel zero on scattering. This appendix establishes the absence of the symmetry obstruction; the cited pairwise-covariant constructions supply the actual nonzero scattering amplitude. There is therefore no ordinary crossing relation between the two four-point state spaces.



\begin{thebibliography}{99}

\bibitem{Dirac:1931kp}
P.~A.~M.~Dirac,
``Quantised singularities in the electromagnetic field,''
\href{https://doi.org/10.1098/rspa.1931.0130}{Proc. Roy. Soc. Lond. A \textbf{133} (1931) no.821, 60-72}.


\bibitem{Terning:2018lsv}
J.~Terning and C.~B.~Verhaaren,
``Dark Monopoles and $SL(2,\mathbb Z)$ Duality,''
JHEP \textbf{12} (2018), 123
\arXiv{1808.09459}{hep-th}.

\bibitem{Nielsen:1973cs}
  H.~B.~Nielsen and P.~Olesen,
  ``Vortex Line Models for Dual Strings,''
  \href{http://dx.doi.org/10.1016/0550-3213(73)90350-7}{Nucl.\ Phys.\ B {\bf 61} (1973) 45}.

\bibitem{Weinberg:1965rz}
S.~Weinberg,
``Photons and gravitons in perturbation theory: Derivation of Maxwell's and Einstein's equations,''
\href{https://journals.aps.org/pr/abstract/10.1103/PhysRev.138.B988}{Phys. Rev. \textbf{138} (1965), B988-B1002}.




\bibitem{Terning:2018udc}
J.~Terning and C.~B.~Verhaaren,
``Resolving the Weinberg Paradox with Topology,''
\href{https://doi.org/10.1007/JHEP03(2019)177}{JHEP \textbf{03} (2019) 177},
\arXiv{1809.05102}{hep-th}.


\bibitem{Zwanziger:1970hk}
D.~Zwanziger,
``Local Lagrangian quantum field theory of electric and magnetic charges,''
\href{https://doi.org/10.1103/PhysRevD.3.880}{Phys. Rev. D \textbf{3} (1971) 880}.

\bibitem{SchwarzSen}
J.~H.~Schwarz and A.~Sen,
``Duality symmetric actions,''
Nucl.\ Phys.\ B {\bf 411} (1994) 35,
\arXivold{hep-th/9304154}.

\bibitem{Terning:2020dzg}
J.~Terning and C.~B.~Verhaaren,
``Spurious Poles in the Scattering of Electric and Magnetic Charges,''
\href{https://doi.org/10.1007/JHEP12(2020)153}{JHEP \textbf{12} (2020) 153},
\arXiv{2010.02232}{hep-th}.


\bibitem{Elvang:2013cua}
H.~Elvang and Y.~t.~Huang,
``Scattering Amplitudes,''
\arXiv{1308.1697}{hep-th}.

\bibitem{Arkani-Hamed:2017jhn}
N.~Arkani-Hamed, T.~C.~Huang and Y.~t.~Huang,
``Scattering amplitudes for all masses and spins,''
\href{https://doi.org/10.1007/JHEP11(2021)070}{JHEP \textbf{11} (2021) 070},
\arXiv{1709.04891}{hep-th}.

\bibitem{Christensen:2022nja}
N.~Christensen, H.~Diaz-Quiroz, B.~Field, J.~Hayward and J.~Miles,
``Challenges with internal photons in constructive QED,''
\href{https://doi.org/10.1016/j.nuclphysb.2023.116278}{Nucl. Phys. B \textbf{993} (2023) 116278},
\arXiv{2209.15018}{hep-ph}.

\bibitem{Lai:2023upa}
H.~Y.~Lai, D.~Liu and J.~Terning,
``The Constructive Method for Massive Particles in QED,''
\href{https://doi.org/10.1007/JHEP06(2024)086}{JHEP \textbf{06} (2024) 086},
\arXiv{2312.11621}{hep-th}.



\bibitem{Csaki:2020inw}
C.~Csaki, S.~Hong, Y.~Shirman, O.~Telem, J.~Terning and M.~Waterbury,
``Scattering amplitudes for monopoles: pairwise little group and pairwise helicity,''
\href{https://doi.org/10.1007/JHEP08(2021)029}{JHEP \textbf{08} (2021) 029},
\arXiv{2009.14213}{hep-th}.


\bibitem{Csaki:2022tvb}
C.~Cs{\'a}ki, Z.~Y.~Dong, O.~Telem, J.~Terning and S.~Yankielowicz,
``Dressed vs. pairwise states, and the geometric phase of monopoles and charges,''
JHEP \textbf{02} (2023) 211,
\arXiv{2209.03369}{hep-th}.



\bibitem{Csaki:2020yei}
C.~Cs\'aki, S.~Hong, Y.~Shirman, O.~Telem and J.~Terning,
``Completing Multiparticle Representations of the Poincar\'e Group,''
\href{https://doi.org/10.1103/PhysRevLett.127.041601}{Phys. Rev. Lett. \textbf{127} (2021) 041601},
\arXiv{2010.13794}{hep-th}.



\bibitem{Ignatiev:1997pm}
A.~Y.~Ignatiev and G.~C.~Joshi,
``Dirac magnetic monopole and the discrete symmetries,'' in
\href{https://doi.org/10.1016/S0960-0779(99)00057-0}{Chaos Solitons Fractals \textbf{11} (2000) 1411},
\arXivold{hep-ph/9710553}.




\bibitem{Britto:2005fq}
R.~Britto, F.~Cachazo, B.~Feng and E.~Witten,
``Direct Proof of the Tree-Level Scattering Amplitude Recursion Relation in Yang-Mills Theory,''
\emph{Phys. Rev. Lett.} \textbf{94} (2005) 181602,
\arXivold{hep-th/0501052}.

\bibitem{ArkaniHamed:2008yf}
N.~Arkani-Hamed and J.~Kaplan,
``On Tree Amplitudes in Gauge Theory and Gravity,''
\emph{JHEP} \textbf{04} (2008) 076,
\arXiv{0801.2385}{hep-th}.

\bibitem{Cheung:2008dn}
C.~Cheung,
``On-Shell Recursion Relations for Generic Theories,''
\emph{JHEP} \textbf{03} (2010) 098,
\arXiv{0808.0504}{hep-th}.

\bibitem{Cohen:2010mi}
T.~Cohen, H.~Elvang and M.~Kiermaier,
``On-shell Constructibility of Tree Amplitudes in General Field Theories,''
\emph{JHEP} \textbf{04} (2011) 053,
\arXiv{1010.0257}{hep-th}.


\bibitem{Benincasa:2011pg}
P.~Benincasa and E.~Conde,
``On the Tree-Level Structure of Scattering Amplitudes of Massless Particles,''
\emph{JHEP} \textbf{11} (2011) 074,
\arXiv{1106.0166}{hep-th}.


\bibitem{Ema:2024rss}
Y.~Ema, T.~Gao, W.~Ke, Z.~Liu, K.~F.~Lyu and I.~Mahbub,
``Momentum shift and on-shell recursion relation for electroweak theory,''
\href{https://doi.org/10.1103/PhysRevD.110.105002}{Phys. Rev. D \textbf{110} (2024) 105002},
\arXiv{2407.14587}{hep-ph}.


\bibitem{Thomson}
J.~J. Thomson, ``On Momentum in the Electric Field,'' \href{https://www.tandfonline.com/doi/abs/10.1080/14786440409463203}{Philos. Mag. 8 (1904) 331}.


\bibitem{Jiang:2020rwz}
M.~Jiang, J.~Shu, M.~L.~Xiao and Y.~H.~Zheng,
``Partial Wave Amplitude Basis and Selection Rules in Effective Field Theories,''
\href{https://doi.org/10.1103/PhysRevLett.126.011601}{Phys. Rev. Lett. \textbf{126} (2021) 011601},
\arXiv{2001.04481}{hep-ph}.

\bibitem{Witten:2003nn}
E.~Witten,
``Perturbative gauge theory as a string theory in twistor space,''
\href{https://doi.org/10.1007/s00220-004-1187-3}{Commun. Math. Phys. \textbf{252} (2004) 189},
\arXiv{0312171}{hep-th}.



\bibitem{Conde:2016izb}
E.~Conde, E.~Joung and K.~Mkrtchyan,
``Spinor-Helicity Three-Point Amplitudes from Local Cubic Interactions,''
\href{https://doi.org/10.1007/JHEP08(2016)040}{JHEP \textbf{08} (2016) 040},
\arXiv{1605.07402}{hep-th}.


\bibitem{Degrande:2025uil}
C.~Degrande, H.~L.~Li and L.~X.~Xu,
``Partial-Wave Unitarity Bounds on Higher-Dimensional Operators from 2-to-$N$ Scattering,''
\arXiv{2511.15524}{hep-ph}.




\bibitem{Moynihan:2020gxj}
N.~Moynihan and J.~Murugan,
``On-shell electric-magnetic duality and the dual graviton,''
\href{https://doi.org/10.1103/PhysRevD.105.066025}{Phys. Rev. D \textbf{105} (2022) 066025},
\arXiv{2002.11085}{hep-th}.



\bibitem{Weinberg:1995mt}
S.~Weinberg,
``The Quantum theory of fields. Vol. 1: Foundations,''
Cambridge University Press, 2005,
ISBN 978-0-521-67053-1, 978-0-511-25204-4


\bibitem{Liu:2022alx}
D.~Liu and Z.~Yin,
``Gauge invariance from on-shell massive amplitudes and tree-level unitarity,''
\href{https://doi.org/10.1103/PhysRevD.106.076003}{Phys. Rev. D \textbf{106} (2022) 076003},
\arXiv{2204.13119}{hep-th}.


\bibitem{Elvang:2015rqa}
H.~Elvang and Y.~t.~Huang,
``Scattering Amplitudes in Gauge Theory and Gravity,''
Cambridge University Press, 2015,
ISBN 978-1-316-19142-2, 978-1-107-06925-1;
``Scattering Amplitudes,''
\arXiv{1308.1697}{hep-th}.



\end{thebibliography}
\end{document}